\documentclass[10pt,letterpaper]{article}
\usepackage[top=0.85in,left=2.75in,footskip=0.75in]{geometry}
\usepackage{float}
\usepackage{amsmath,amssymb}

\usepackage{changepage}

\usepackage[utf8x]{inputenc}

\usepackage{textcomp,marvosym}

\usepackage{cite}

\usepackage{nameref,hyperref}

\usepackage[right]{lineno}

\usepackage{microtype}
\DisableLigatures[f]{encoding = *, family = * }

\usepackage[table]{xcolor}

\usepackage{array}

\usepackage[symbol]{footmisc}

\newcolumntype{+}{!{\vrule width 2pt}}

\newlength\savedwidth

\raggedright
\usepackage[aboveskip=1pt,labelfont=bf,labelsep=period,justification=raggedright,singlelinecheck=off]{caption}

\makeatletter
\renewcommand{\@biblabel}[1]{\quad#1.}
\makeatother

\date{}

\usepackage{lastpage,fancyhdr,graphicx}
\usepackage{epstopdf}
\fancyheadoffset[L]{2.25in}
\fancyfootoffset[L]{2.25in}

\definecolor{myblue}{rgb}{0,0,0}

\begin{document}
\vspace*{0.2in}

\begin{flushleft}
{\Large
\textbf\newline{Slow diffusive dynamics in a chaotic balanced neural network \footnote{PLoS Computational Biology, in press (2017)}} 
}
\newline
\\
Nimrod Shaham\textsuperscript{1},
Yoram Burak\textsuperscript{1,2}

\bigskip
\textbf{1} Racah Institute of Physics
\\
\textbf{2} Edmond and Lily Safra Center for Brain Sciences
\\
 The Hebrew University of Jerusalem, Jerusalem, Israel
\\
\bigskip

%
%



\end{flushleft}
\section*{Abstract}
It has been proposed that neural noise in the cortex arises from chaotic dynamics in the
balanced state: in this model of cortical dynamics, 
the excitatory and inhibitory inputs to each neuron approximately cancel, 
and activity is driven by fluctuations of the synaptic inputs around their mean. 
It remains unclear whether neural networks in the balanced state can perform tasks that are 
highly sensitive to noise, such as storage of continuous parameters in working memory, while also accounting for the irregular behavior of single neurons. 
Here we show that continuous parameter working memory can be maintained in the balanced state, in a neural circuit with
a simple network architecture.
We show analytically that in the limit of an infinite network, the dynamics generated by this architecture
are characterized by 
a continuous set of steady balanced states, allowing for the indefinite storage of a continuous parameter. 
In finite networks, we show that the chaotic noise drives diffusive motion along the approximate attractor, which 
gradually degrades the stored memory.
We analyze the dynamics and show that 
the slow diffusive motion induces slowly decaying temporal cross correlations in the activity, 
which differ substantially from those previously described in the balanced state. 
We calculate the diffusivity, 
and show that it is inversely proportional to the system size.
For large enough (but realistic) neural population sizes, and with suitable tuning of the network connections, the proposed
balanced network can sustain continuous parameter values in memory
over time scales larger by several orders of magnitude than the single neuron time scale.


\section*{Introduction}
The consequences of irregular activity in the brain, and the mechanisms responsible for its emergence, are topics of fundamental interest in the study of brain function and dynamics. In theoretical models of brain activity, the irregular dynamics observed in neuronal activity are often modeled as arising from noisy inputs or from intrinsic noise in the dynamics of single neurons. However, theoretical and experimental works have suggested that explanations based on sources of noise in intrinsic neural dynamics are insufficient to account for the stochastic nature of activity in the cortex 
\cite{mainen1995reliability,holt1996comparison,van1996chaos,shadlen1998variable}. 
An alternative proposal is that noise in the cortex arises primarily from chaotic dynamics at the network level 
\cite{sompolinsky1988chaos,van1996chaos,shadlen1998variable,wolf2010chaos}.
A central result in the field is that simple neural circuits with recurrent random connectivity can settle, under a broad range of conditions, into a fixed point called the balanced state 
\cite{van1996chaos, van_vreeswijk_chaotic_1998, van2005houches, renart2010asynchronous}: 
in this state the mean excitatory drive to each neuron nearly balances the mean inhibitory drive, and neural activity is driven by fluctuations in the excitatory and inhibitory inputs. The overall dynamics are chaotic, resulting in an apparent stochasticity in the activity of single neurons, which can exist even in the absence of any sources of random noise intrinsic to the dynamics of single neurons and synapses. 

It remains unclear which computational functions in the brain are compatible with the architecture of the balanced network model, since this model assumes random, unstructured connectivity in its rudimentary form. The possibility that functional circuits in the brain are in a balanced state raises another important question: does the apparent stochasticity of single neurons in this state have similar consequences on brain function as would arise from stochasticity which is truly intrinsic to the neural and synaptic dynamics? 

Here we explore the effects of chaotic noise on continuous parameter working memory.
This task is particularly 
sensitive to noise, yet neurons in cortical areas involved in the maintenance of continuous parameter working memory have been shown  
to fire irregularly during task performance \cite{compte2003temporally,wimmer2014bump}.
Attractor dynamics are often put forward as a mechanism for the persistent neural activity underlying this task. Continuous attractor networks are dynamically characterized by a continuous manifold of semi-stable steady states, which make it possible to memorize parameters with a continuous range of values, such as an angle or a position 
\cite{robinson_integrating_1989, tsodyks1995rapid, zhang1996representation, seung_how_1996,compte2000synaptic, renart2003robust ,barak_working_2014}. In such networks, noise in neural or synaptic activity can cause diffusion along the manifold of steady states, leading to gradual degradation of the stored memory 
\cite{burak_fundamental_2012,kilpatrick2013optimizing,bays2014noise}. However,
irregular activity in the balanced state does not arise from mechanisms intrinsic to neurons or synapses, but rather from chaotic dynamics, and its consequences for continuous parameter working memory are largely unexplored. For this reason we addressed two questions. First,
we asked whether a neural network can possess a continuum of balanced stable states. Second, we investigated how, in this scenario, chaotic noise 
would affect information maintenance. 

\subsubsection*{Persistence in balanced networks}

The question of whether balanced networks can produce persistent activity has attracted considerable interest in recent years. Several works explored architectures which give rise to slow dynamics in balanced networks, characterized by the coexistence of multiple discrete balanced states \cite{renart_mean-driven_2007}. In several recent works multi-stability resulted from the existence of clustered connectivity, 
and slow transitions were observed between the discrete semi-stable states
\cite{litwin2012slow,rosenbaum2014balanced,stern2014dynamics}. Other works 
\cite{van2005houches, roudi2007balanced} demonstrated that a discrete set of semi-stable states can
be embedded in a balanced neural network, using a similar construction as employed in the classical Hopfield model of associative memory 
\cite{hopfield1982neural}. 

A few works have addressed the possibility that balanced neural networks may generate slow persistent activity over a continuous manifold.
Such dynamics were demonstrated in simulations of neural networks that included short-term synaptic plasticity
\cite{hansel_short-term_2013}, or a derivative-feedback mechanism \cite{lim_balanced_2013,lim_balanced_2014}. 
Previous works have not demonstrated   
the existence of a continuum of steady states in a balanced neural network analytically, and it has remained unclear whether such a continuum can be obtained without evoking additional mechanisms (such as 
short-term synaptic plasticity, or derivative-feedback). In addition, the influence of the chaotic dynamics on the persistence of stored memory has not been analyzed. These questions are addressed in the present work. 

Below, we identify an architecture in which slow dynamics are attainable in a simple form of a balanced neural network. We prove analytically the existence of a continuous attractor in our model in the large population limit. In finite networks, we show that the chaotic noise drives diffusive motion along the attractor -- leading, among other consequences, to slowly decaying spike cross-correlations. We show that the diffusivity scales inversely with the system size, as predicted previously for continuous attractor networks with intrinsic sources of neuronal stochasticity. With a reasonable number of neurons and suitable tuning, our model network 
exhibits slow dynamics over a continuous manifold of semi-stable states, 
while exhibiting single neural dynamics which appear stochastic, as observed in cortical circuits.


\section*{Results}

\subsection*{Reciprocal inhibition between two balanced networks}

Our neural network model is based on the classical balanced network model presented in Refs.~\cite{van1996chaos,van_vreeswijk_chaotic_1998, van2005houches}. This model consists of two distinct populations of binary neurons, one inhibitory and the other excitatory. The recurrent connectivity is random and sparse, with a probability $K/N$ for a connection, where $N$ is the population size (assumed for simplicity to be the same in both populations), $K$ is the average number of connections per neuron from each population, and the connection strength is $\sim 1/\sqrt{K}$. 
For $1\ll K \ll N$ and over a wide range of parameters, the mean population activity settles to a fixed point 
(the \textit{balanced state}) where on average the total excitation received by each neuron is approximately canceled by the total inhibition (to leading order in $1/\sqrt{K}$),
and the neural dynamics are driven by the fluctuations in the input. The single neuron activity appears noisy, neither of the populations is fully activated or deactivated, and the overall network state is chaotic. 
\\
\begin{figure}[tb]
\centering
\medskip
\includegraphics[scale=1.0]{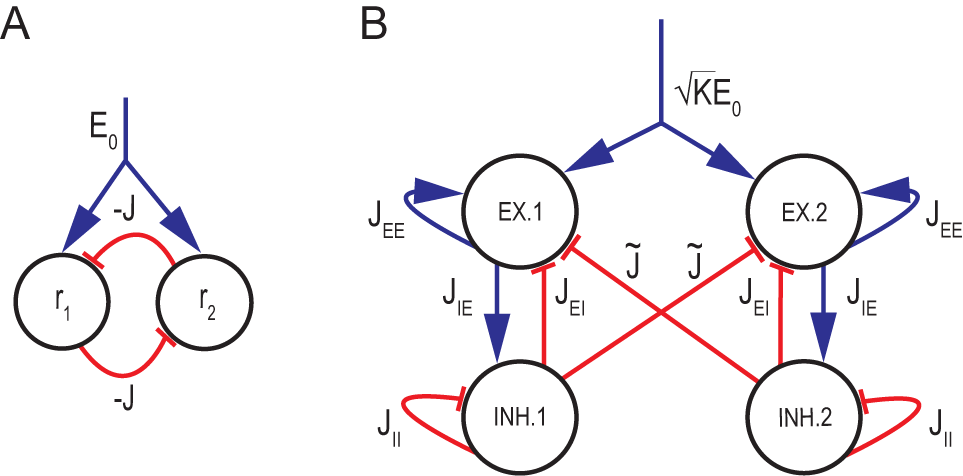} 
\vspace{2mm}
\caption{\textbf{Network architecture and parameters}. \textbf{A} Two neural populations with rates $r_1,\; r_2$ inhibit each other with synaptic efficacies $-J$. \textbf{B} Two coupled balanced subnetworks, each consisting of an excitatory and inhibitory population. Excitatory (inhibitory)
connections are represented in blue (red).
Mutual inhibition is generated by all-to-all connections of strength 
$-\tilde{J}\sqrt{K}/N$ from each inhibitory population to the excitatory population of the other subnetwork. 
As in \cite{van_vreeswijk_chaotic_1998},
connections within each subnetwork are random with a connection probability $K/N$,  $1\ll K \ll N$. Connection strengths are:  $ J_{EE}/\sqrt{K},\:   J_{IE}/\sqrt{K}, \: J_{EI}/\sqrt{K}$ and $ J_{II}/\sqrt{K} $ according to the identity of the participating neurons. Without loss of generality, we chose $J_{EE}=J_{IE}=1$ and define $J_{EI}\equiv -J_E, J_{II}\equiv -J_I$. An excitatory input 
$\sqrt{K} E_0$ is fed into both excitatory populations}
%
\label{fig:arch}
\end{figure}

Despite the nonlinearities involved in the dynamics of each neuron, the population averaged activities in the balanced state are linear functions of the external input \cite{van1996chaos,van_vreeswijk_chaotic_1998}. We 
exploit this linearity to build a simple system of two balanced networks projecting to each other. The intuition comes from 
a simple model of a continuous attractor neural network consisting of linear neurons arranged in two populations that mutually inhibit each other, Fig.~\ref{fig:arch}A. 
The linear rate dynamics of this system are given by: 
\begin{equation}
    \tau \dot{\boldsymbol{r}}=-\boldsymbol{r}+W\boldsymbol{r}+\boldsymbol{E} \, ,
    \label{simple_dynamics}
\end{equation}
where $\boldsymbol{E}=[E_0,E_0],\; E_0>0$ is an external input and
\begin{equation}
W= \left( \begin{array}{cc}
0 & -J  \\
-J & 0 \end{array} \right)\, .
\label{simple_W}
\end{equation}
For $J = 1$ the system has a vanishing eigenvalue, and the fixed points form a continuous line: $r_1+r_2=E_0$. 

The simple neural architecture of Fig.~\ref{fig:arch}A 
was used as a basis for modeling the dynamics of neural circuits responsible for memory and decision making in the prefrontal cortex
\cite{cannon1983proposed,wang2002probabilistic,machens2005flexible,deco2006decision,Doiron2012variability}.  
In our model, a balanced subnetwork replaces each of the populations of Fig.~\ref{fig:arch}A, 
and the inhibitory population in each subnetwork projects to the excitatory population of the other subnetwork, Fig.~\ref{fig:arch}B
and \textit{Methods}. Thus, the model consists of two reciprocally inhibiting balanced neural populations. 

{\color{myblue} We consider first a scenario in which the inhibitory connectivity between the two sub-networks is all-to-all. Therefore, the network includes a combination of strong, random synapses within each sub-network and highly structured, weak synapses between the two sub-networks. This scenario lends itself to analytical treatment of finite $N$ effects (see below). Later on, we present results also for an alternative scenario, in which the connections between the two-subnetworks are sparse, random, and strong (\textit{Additional randomness in connectivity and inputs}).} 

\subsection*{Continuum of balanced states}

We first examine whether the two-subnetwork architecture can give rise to a continuum of balanced states.
The parameters of the network connectivity in our model
are summarized in Fig.~\ref{fig:arch} and in \textit{Methods}. The mutual inhibition between 
the subnetworks is assumed to be all to all, and the interaction strength
is scaled such that the total inhibitory input to each neuron, coming from the opposing subnetwork
scales in proportion to $\sqrt{K}$.

Similar to the case of a single balanced network \cite{van_vreeswijk_chaotic_1998}, 
the mean field dynamics of the population averaged activities for $N \rightarrow \infty$ and $K \gg 1$ are given by:
\begin{equation}
\tau_i \dot{m}_i=-m_i+H(-u_i/\sqrt{\alpha_i})\,,
\label{full_diff_sys}
\end{equation}
where $m_i(t)=1/N \sum_{k=1}^N \sigma_i^k(t)$ [$i=1$ (2) for the excitatory
(inhibitory) population of the first subnetwork, and similarly $i=3,4$ in the second subnetwork], $\sigma_i^k(t)$ is the state of neuron $k$ in population $i$ at time $t$, $H(x)$ is the complementary error function,
and $u_i$ ($\alpha_i$) is the mean (variance) of the input to a neuron in population $i$, averaged over
the population and over the random connectivity (\textit{Methods}).
{\color{myblue} This equation is an approximation which becomes
exact in the limit $K \rightarrow \infty$}.

To check whether there exist parameters for which the system has a continuum of balanced states, it is convenient to write the 
steady state equations of the above dynamics, while making use of the
assumption that $K$ is large. In the limit $K \rightarrow \infty$
these equations become linear (\textit{Methods}):
\begin{equation}
\begin{array}{ccl}
	m_1-J_E m_2-\tilde{J} m_4 +E_0 &=& 0 \,,\\
	m_1-J_I m_2 &=& 0 \,,\\
	m_3-J_E m_4-\tilde{J} m_2 +E_0 &=& 0 \,,\\
	m_3-J_I m_4 &=& 0 \,.
\end{array}
\label{full_sys_lin}
\end{equation}
By choosing the interaction strength between the two subnetworks to be $\tilde{J}=J_E-J_I$, this system becomes singular, and has a continuum of solutions arranged on a line in the mean activities space, which represent a continuum of stable balanced states. 

\subsubsection*{Finite $K$}

When $K$ is {\color{myblue} large and finite}, but still in the idealized limit of infinite $N$, 
the population dynamics remain deterministic, and are 
{\color{myblue} approximately} described by Eq.~\ref{full_diff_sys}, but 
the nullclines corresponding to the steady state equations (Eq.~\ref{full_sys}) are now nonlinear. 
Therefore, a precise continuum of steady states cannot be established. 
However, if the nonlinear nullclines are close to each other, slow dynamics are attainable in a specific direction of the mean activity space. 
To make this statement more precise,
we first note that the steady state equations always have a symmetric solution in which $m_1=m_3$ and $m_2=m_4$. 
If in addition, at this symmetric point,
the slopes of the nullclines are identical 
($\partial m_3 / \partial m_1=-1$), there is a vanishing eigenvalue of the linearized mean field dynamics 
(Eq.~\ref{full_diff_sys})
around this point (\textit{Methods}). 

\begin{figure}
\centering
\medskip
\includegraphics[scale=1.0]{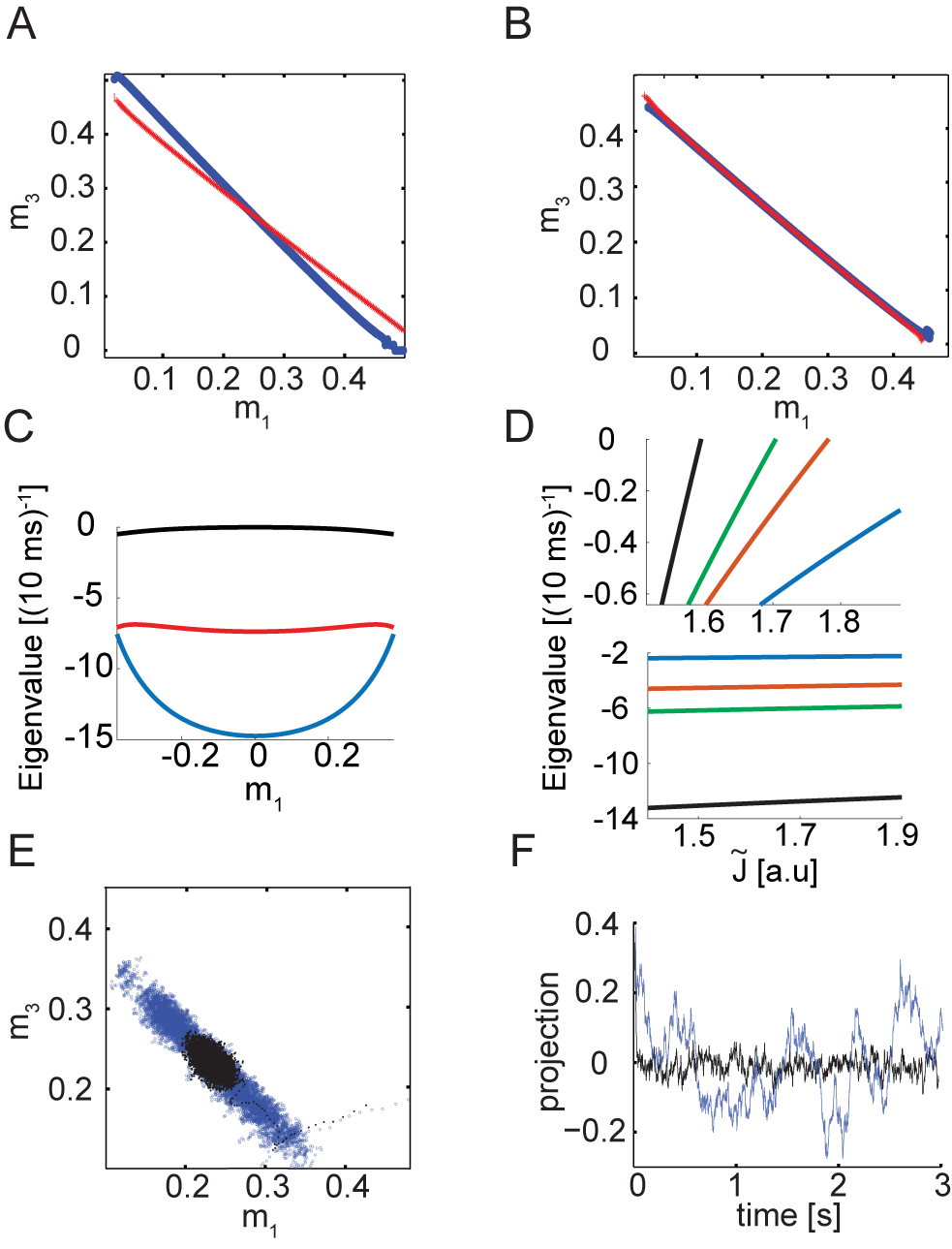} 
\vspace{2mm}
\caption{\textbf{Dynamics in the $\boldsymbol{N\rightarrow \infty}$ limit}. \textbf{A}  Projections of the nullclines $\dot{m}_1=0$ (blue) and $\dot{m}_3=0$ (red) on the $m_1 - m_3$ plane, based on Eq.~\ref{full_sys}.
Here
$K=1000,\; J_E=4,\; J_I=2.5,\; \tilde{J}= 1.5,\; \tau_E=10\; \text{ms},\; \tau_I=8\; \text{ms}, \; \text{and} \; E_0=0.3$.
 The same values of $J_E, J_I, E_0, \tau_E$, and $\tau_I$ are used throughout the manuscript.
\textbf{B} Same as A, except that here $\tilde{J}\simeq 1.7$, tuned to achieve a singular Jacobian at the symmetric point.
\textbf{C} Real part of the four eigenvalues of the Jacobian evaluated at different points along the approximate attractor (here parametrized by the value of $m_1$).
Note that there are two complex conjugate eigenvalues, and their real parts overlap (red curve).
\textbf{D} Top: the eigenvalue closest to zero as a function of the mutual inhibition strength. 
Different colors correspond to different values of $K$:  $5000$ (black), $1000$  (green), $500$ (red) and $100$ (blue). 
All other parameters are as in A. Bottom: real part of the eigenvalue next to closest to zero as a function of 
the mutual inhibition strength.
\textbf{E} Integration of Eq.~\ref{full_diff_sys} with injected uncorrelated Gaussian noise with 
 $\sigma=10^{-2} \; \sqrt{10 \text{msec}}$  (Eqs.~\ref{noise_dyn}-\ref{white_noise}), $\tilde{J}=1.5$ (black), $\tilde{J}\simeq 1.7$ (blue).
\textbf{F} Dynamics of the projection along the special direction (parameters and colors as in E).
}
\label{fig:finite_K}
\end{figure}

Under these conditions, the smallest eigenvalue 
of the linearized dynamics is expected to be small also in the vicinity of the symmetric point. In fact, even for moderately large values of $K$ ($K=1000$), the two nullclines nearly overlap over a large range of $m_1$ and $m_3$, Fig.~\ref{fig:finite_K}A and \ref{fig:finite_K}B. 
Figure~\ref{fig:finite_K}C demonstrates that in this case there is an eigenvalue close to zero 
within a wide range of locations along the approximate attractor, and therefore the dynamics are 
slow at any position along this range.
Below, we denote by $\lambda$ the eigenvalue closest to zero of the linearized dynamics, evaluated at the 
symmetric fixed point. 

As  observed in other continuous attractor neural network models, 
the slow dynamics are sensitive to the
tuning of the recurrent connectivity 
\cite{robinson_integrating_1989,seung_how_1996,machens2005flexible,lim_balanced_2013} (see also \textit{Discussion}). 
This sensitivity is quantified by the dependence of
$\lambda$ on the coupling strength between the two subnetworks.
Figure~\ref{fig:finite_K}D (top panel) shows how $\lambda$
depends on the mutual inhibition strength $\tilde{J}$ and on $K$: $\lambda$ is linear in $\tilde{J}$, and proportional to $\sqrt{K}$. For $K=1000$, $\tilde{J}$ must be tuned to a precision of $\sim 0.1\%$ to achieve a time scale $\lambda^{-1}$ of several seconds, when the intrinsic time scale $\tau$
is $10$ ms. The real part of the next eigenvalue is negative, proportional to $\sqrt{K}$, and is weakly dependent on $\tilde{J}$ (Fig.~\ref{fig:finite_K}D, bottom panel). 

We find that the approximate line attractor is stable to small
perturbations over a wide range of parameters. This is verified by observing that when linearizing the population dynamics 
(Eq.~\ref{full_diff_sys}) around positions
along the approximate attractor, the real parts of the four eigenvalues are negative, and one of them is close to zero, reflecting
the slow dynamics along the approximate attractor -- as 
demonstrated in Fig.~\ref{fig:finite_K}C.

 As an illustration of the existence of a direction in mean activity space, along which the dynamics are slow, we numerically 
 solved the mean field differential equations in the limit of infinite $N$ and $K=1000$, with injected white noise. Figure \ref{fig:finite_K}(E-F) shows that the 
 resulting mean activities trace a line in the mean activities space (E) and the dynamics along the line are slow (F).  


\subsection*{Diffusive dynamics in finite size networks}

Next, we consider the realistic situation in which $N$ is finite in the two-subnetwork model, 
while still requiring that $N \gg K \gg 1$. Instead of adding noise to the dynamics of each neuron, 
we ask whether the 
chaotic dynamics are sufficient to drive diffusive motion along the approximate attractor. This
question is motivated by the fact that diffusive dynamics are
observed in model neural networks of intrinsically noisy neurons, with a finite number of neurons \cite{burak_fundamental_2012}.
In addition, this question is motivated by evidence of diffusive dynamics underlying continuous parameter
working tasks -- as observed both in the
behavioral data and in its neural correlates in the prefrontal cortex
\cite{funahashi1989mnemonic,compte2003temporally,wimmer2014bump}. 

Since the population dynamics are no longer given by Eq.~\ref{full_diff_sys}, we 
performed large scale numerical simulations
of networks with $N$ ranging between $10^4$ to $15 \times 10^4$ (additional details on the simulations can be found in \textit{Methods}).
To simplify the analysis and the presentation,
we chose the random weights within each subnetwork such that they precisely mirrored each 
other, which ensured that the fixed point would be symmetric ($m_1=m_3$ and $m_2=m_4$). If, alternatively, the connections in each subnetwork are chosen independently, the fixed point deviates slightly from this symmetry plane (this deviation approaches zero for infinite networks). However, all the results described below remain qualitatively valid
{\color{myblue} (see below, \textit{Additional randomness in connectivity and inputs})}.

\begin{figure}
\centering
\medskip
\includegraphics[scale=1.0]{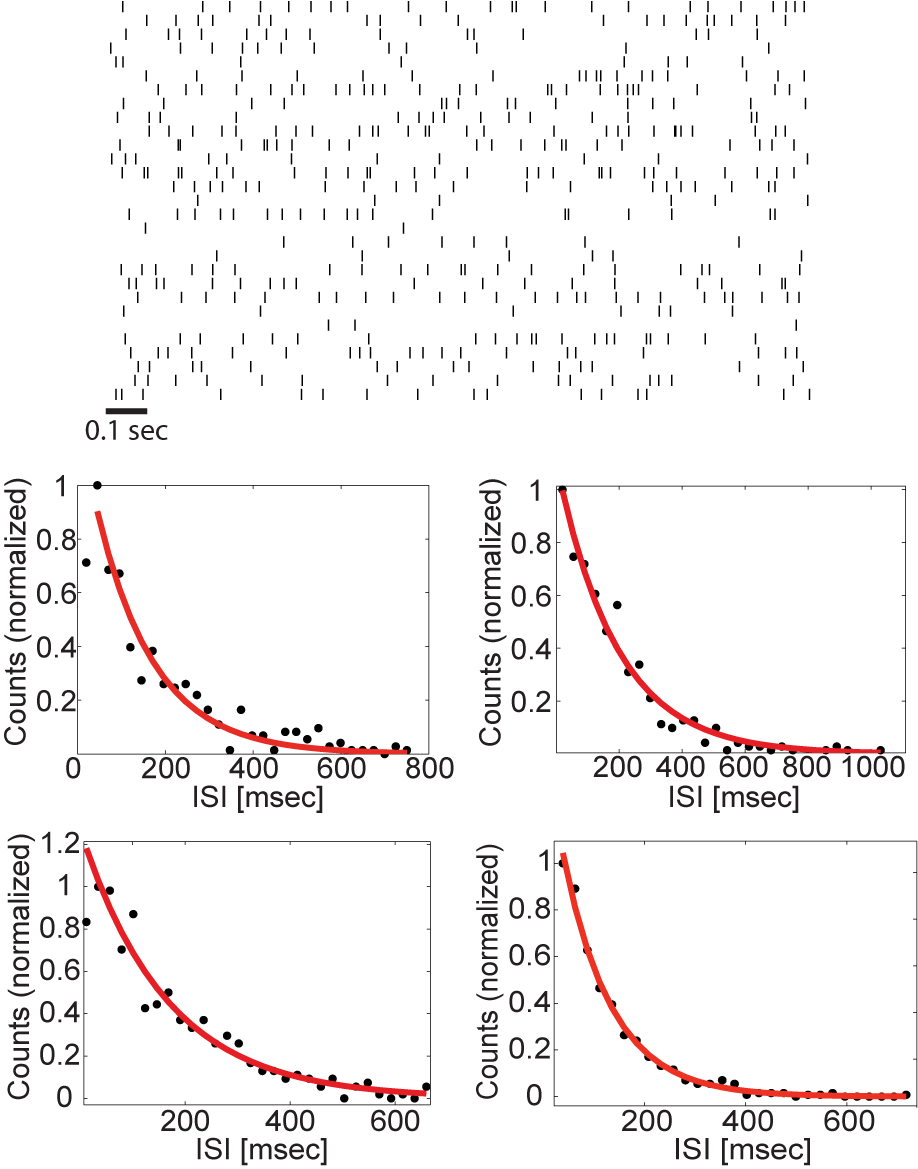} 
\vspace{2mm}
\caption{\textbf{Single neuron statistics}. Top: Raster plot of 30 neurons 
from one of the excitatory populations in resting state. Here a spike is defined as a transition from an `off' state to an `on' state. Bottom: inter-spike interval distribution of four representative neurons from one of the excitatory populations. A fit to an exponential function is shown as a solid red line.} 
\label{fig:ISI}
\end{figure}

\begin{figure}[H]
\centering
\medskip
\includegraphics[scale=1.0]{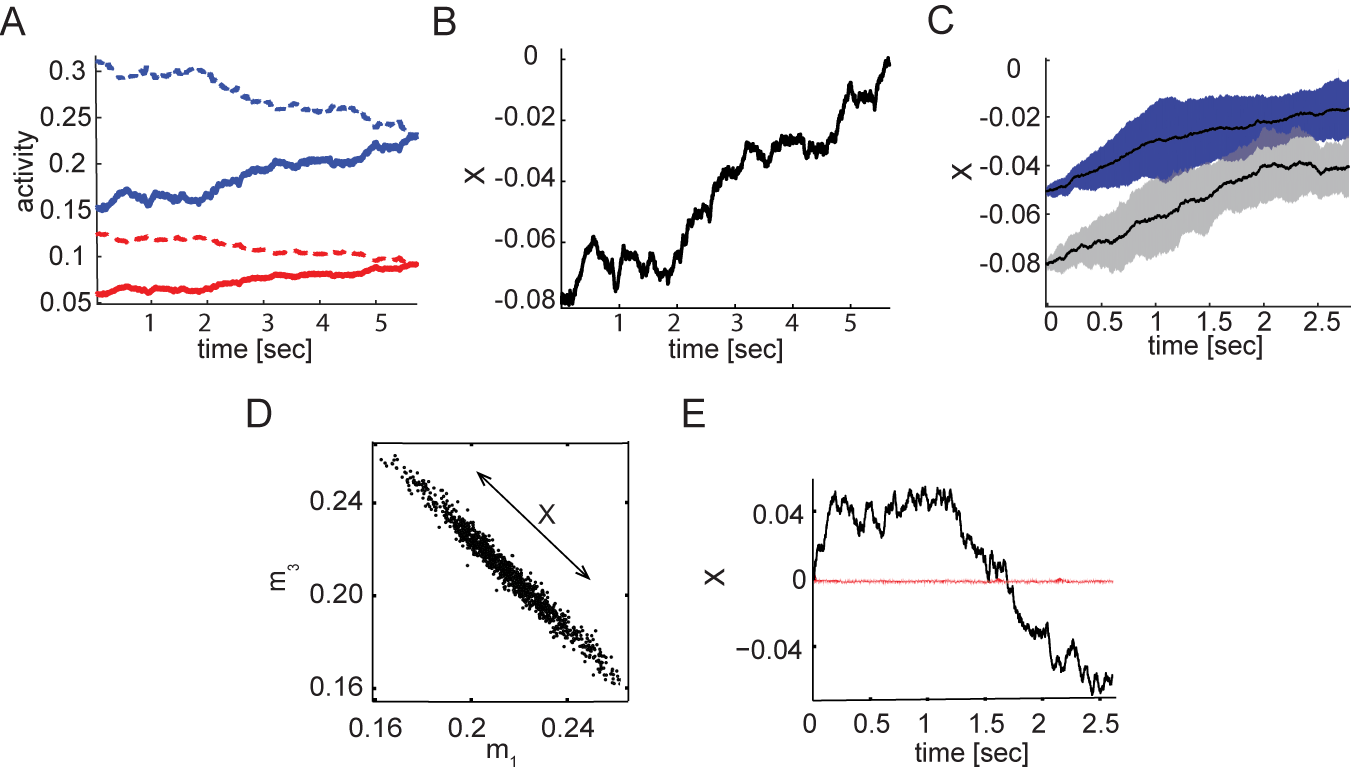} 
\vspace{2mm}
\caption{\textbf{Dynamics of finite $N$ Networks}. \textbf{A} Mean activities of the four populations after initialization at a specific state along the approximate attractor. Blue (red) traces are used for the excitatory (inhibitory) populations. Dashed and solid lines are used to distinguish between the two subnetworks. \textbf{B} Projection along the approximate attractor, shown for the same simulation as in A. 
\textcolor{myblue}{\textbf{C} Projected dynamics for two initial conditions: $X(0)=0.05$ (blue shaded area) and  $X(0)=0.08$ (grey shaded area). Black lines are averages over 50 trials. Shaded areas 
represent the standard deviation measured over 50 independent trials.}  
\textbf{D} Projection of the mean activities on the $m_1$-$m_3$ plane in resting state activity. \textbf{E} Dynamics of the projection $X$ along the special direction (black), and a projection on a perpendicular direction (red) in the same simulation as in D.
 In A,B and C $N=1.5\times 10^5$, $K=500$ and $\tilde{J}\simeq 1.77 $. In D and E $N= 10^5$,  $K=1000$ and  $\tilde{J}\simeq 1.69 $. Other parameters are as in Fig.~\ref{fig:finite_K}. 
 } 
\label{fig:finite_dyn}
\end{figure}

The neural activity observed in our simulations is irregular and individual neurons approximately exhibit exponential ISI distributions similar to those observed in the two population case, although their dynamics are deterministic (Fig.~\ref{fig:ISI}). 
To test whether the network can perform short term memory tasks, we initiated 
the population activities such that the network state was close to some point along the 
approximate line attractor. 
Figure \ref{fig:finite_dyn}A shows the resulting dynamics of the four populations: the activities persisted
for a few seconds before decaying towards the symmetric fixed point. Figure \ref{fig:finite_dyn}B shows the projection along the slow direction, $X(t)$
(defined in Eq.~\ref{proj_def}), again revealing the slow decay of the initial state.
\textcolor{myblue}{
Figure \ref{fig:finite_dyn}C shows statistics of trajectories that start from two initial positions along the approximate attractor, when $\tilde J$ is tuned to achieve $\lambda^{-1} \simeq 9\,s$. The state of the network enables discrimination between the two conditions over a time scale of several seconds. The ability to do so with high confidence 
is influenced both by $\lambda$ and the stochasticity of the motion, which we characterize in the following section (see also \textit{Discussion}). S1~Fig shows the mean square displacement (MSD) from the starting
point for the same dataset, averaged over all trials.}

Long after initialization, the population activities fluctuate
around the 
symmetric fixed point, along a line corresponding to the approximate attractor: a projection on the $m_1 - m_3$ plane is shown in
Fig.~\ref{fig:finite_dyn}D.
Fig.~\ref{fig:finite_dyn}E demonstrates that $X(t)$ exhibits slow diffusive dynamics. 
To demonstrate that the dynamics 
are effectively one dimensional, a projection on a perpendicular direction is shown as well. 

\subsubsection*{Statistics of the diffusive motion}

\begin{figure}
\centering
\medskip
\includegraphics[scale=1.0]{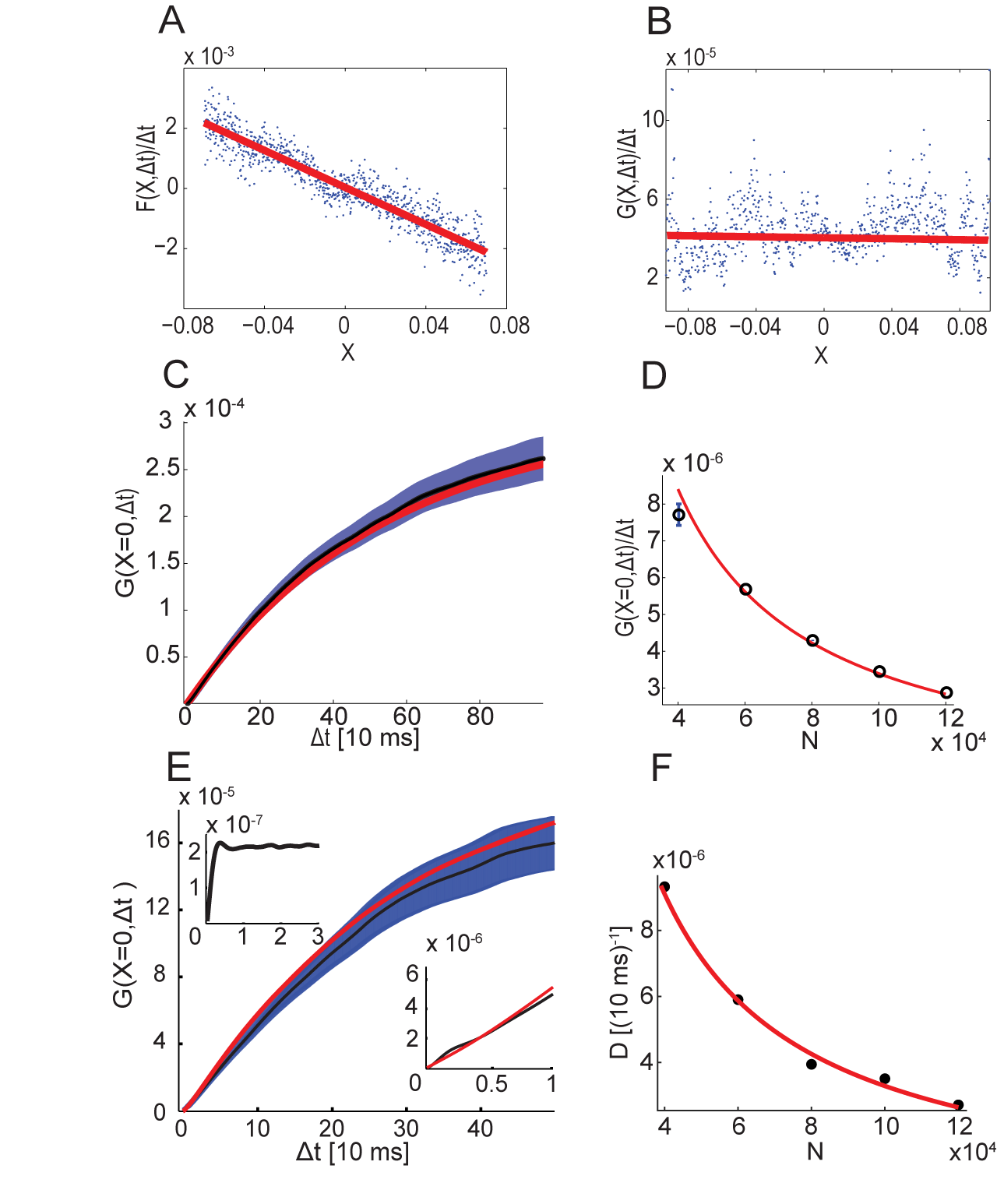}
\vspace{2mm}
\caption{\textbf{Statistics of motion along the approximate attractor}. \textbf{A}-\textbf{B} Statistics of motion along the attractor over short time scales: mean rate of drift 
$F(X,\Delta t \rightarrow 0)/\Delta t$ (A), and random diffusion
$G(X,\Delta t \rightarrow 0)/\Delta t$ (B), measured in numerical simulations as a function of the location $X$
along the approximate attractor. A linear fit is shown in red in both panels.
Here $N=30000$, $K=1000$, $\Delta t\simeq 3$ ms. \textbf{C} Numerical measurement  of $G(X=0,\Delta t)$ (Black trace. The shaded blue area designates the standard deviation of the mean) and a fit to the statistics of an OU process (red).
\textbf{D} Measurement of $G(X,\Delta t\rightarrow 0$) 
(black rings, error bars are smaller than the rings if not shown),
compared with the analytical expression, Eq.~\ref{analytical_short} (red trace). Here $N=1.2\times 10^5$.   \textbf{E}  Measurements of $G(X,\Delta t)$ from
simulations (Black trace. The shaded blue area designates the standard deviation of the mean, same as in C), compared with the semi-analytical approximation (red), Eq.~\ref{semi_analytical} 
($N=1.2\times 10^5$). Lower inset: zoom-in on $\Delta t \leq \tau$. Upper inset: 
measurement of $G(X=0,\Delta t)$ from a single balanced network. Here we chose $X=m_1(t)- \langle m_1(t)  \rangle_t$. \textbf{F} Diffusion coefficient (extracted from a fit to an OU process), 
shown as a function of $N$. Symbols: simulations, red trace: fit to $\sim 1/N$ dependence.
 }
\label{fig:diff_stat}
\end{figure}

From here on we focus on the dynamics of $X(t)$, the projected position along the approximate attractor.  
The dynamics of 
$X$ can be characterized by the two moments:
\begin{eqnarray}
F(X,\Delta t) & \equiv & \left\langle \left. X(t+\Delta t)-X(t) \right| X(t)=X \right\rangle_t \,,
\label{drift1} \\
G(X,\Delta t) & \equiv & \langle \left. \left[X(t+\Delta t)-X(t)\right]^2 \right| X(t)=X \rangle_t \,.
\label{diffusion1}
\end{eqnarray}
The first moment $F$ characterizes the systematic component of drift along the approximate
attractor, and the second moment $G$ characterizes the random, diffusive component of the motion. Both moments may depend, in
general, on the position $X$ along the approximate attractor. Figures \ref{fig:diff_stat}(A-B)
show measurements from simulations of $F(X,\Delta t)$ and $G(X,\Delta t)$ in the limit of small $\Delta t$,
at various locations $X$.

For small $\Delta t$ and near the fixed point, we expect
$F(X,\Delta t) \simeq -\lambda X \Delta t$ with constant $\lambda$, 
where in the limit of large $N$ and $K$, $\lambda$ becomes equal to the smallest eigenvalue of the deterministic linearized dynamics
at the symmetric fixed point.
In fact, 
this relation holds to a very good approximation over a wide range of positions along
the approximate attractor (Fig.~\ref{fig:diff_stat}A). 

The moment $G(X,\Delta t)$ (Eq.~\ref{diffusion1}) characterizes the diffusion along the approximate line attractor,
and is the focus of our analysis in the rest of this section,
since it quantifies the random aspect of the dynamics, 
driven by the chaotic noise.
Figure \ref{fig:diff_stat}B
shows measurements of this quantity from simulations (for small $\Delta t$), over a wide 
range of positions along the approximate attractor. {\color{myblue} Note that for the parameters we use and our choice for the parametrization of $X$ (\textit{Methods}), the range of $X$ is approximately $[-0.2,0.2]$.} Figure \ref{fig:diff_stat}C
shows measurements of $G(X,\Delta t)$ for $X = 0$, over a wide range of time intervals.
 
\subsubsection*{Short time scales}

Our main interest lies in the diffusive motion over long time scales compared to $\tau$. 
However, we consider 
first the diffusive motion over short time scales, quantified by $G(X,\Delta t)$ for $\Delta t \lesssim \tau$, since
in this case the behavior of $G$ can be 
expressed exactly in terms of the
averaged autocorrelation function, $q_j(\Delta t) \equiv 1/N\sum_{i=1}^{N}\langle \sigma_i^j(t+\Delta t) \sigma_i^j(t) \rangle$ (\textit{Methods}).
Using the mean field theory, it is possible to derive a differential equation for 
$q(t)$ \cite{van_vreeswijk_chaotic_1998}, 
which can be solved numerically. Using the numerical solution, 
we obtained a prediction for $G(X,\Delta t)$ which is 
in excellent agreement with measurements from numerical simulations, Fig.~\ref{fig:diff_stat}D. Note that there are no fitting parameters in this calculation.

This analysis leads to two conclusions, which are important for the analysis that follows below: 
first, $G$ is proportional to $\Delta t$ for small $\Delta t$. 
Second, $G$ is inversely proportional to $N$, the size of the neural populations. 
A similar derivation can be applied also to the single balanced network discussed in
\cite{van_vreeswijk_chaotic_1998},
for $\Delta t \ll \tau$
(upper inset in Fig.~\ref{fig:diff_stat}E).

In addition, we note that $G(X,\Delta t \rightarrow 0)/\Delta t$ is approximately constant along the approximate attractor, 
as seen in Fig.~\ref{fig:diff_stat}B.
Therefore, in most of the numerical results below we focus on $G$ near the symmetric fixed point
($X = 0$).
%

\subsubsection*{Diffusion over arbitrary time scales}
On time scales larger than $\tau$, the behavior of the two coupled balanced subnetworks differs dramatically from that 
of the single balanced network:
in the single balanced network
$G$ saturates for $\Delta t \gtrsim \tau$
(Fig.~\ref{fig:diff_stat}E, upper inset),
whereas in the two coupled balanced subnetworks
$G$ continues to increase as a function of $\Delta t$, 
up to $\Delta t$ of order $\lambda^{-1}$
(Fig. \ref{fig:diff_stat}E, main plot). 
Thus, the diffusive motion generates correlated activity over time scales much longer than $\tau$.
Because the chaotic noise itself is uncorrelated on time scales longer than $\tau$ (as shown more precisely below), 
and
since $\lambda$ is approximately constant along the approximate attractor, 
we may expect the motion to approximately follow the statistics of an Ornstein–Uhlenbeck (OU) process.
This approximation provides a good fit to the dynamics, Fig.~\ref{fig:diff_stat}C, as
expected. 
This made it possible to extract a diffusion coefficient $D$ from the simulations which characterizes the random
motion on time scales $\tau \lesssim \Delta t \lesssim \lambda^{-1}$. {\color{myblue} Furthermore, since $D$ and $\lambda$ are approximately constant over a wide range of positions along the approximate attractor (Figs.~\ref{fig:diff_stat}[A-B]), the approximation as an OU process provides a precise and compact description of the trajectory statistics, from which the performance of the network in retention of memory can be deduced (see \textit{Discussion} and Figs.~\ref{fig:diff_stat}C, S1~Fig, S2D~Fig, S3E~Fig and S4[B,C]~Fig)}. 

According to Eq.~\ref{analytical_short} (\textit{Methods}), fluctuations in the mean activity scale as $1/N$, but this equation is valid only for time scales smaller than $\tau$, whereas the diffusion coefficient $D$ characterizes fluctuations on longer time scales.
Figure \ref{fig:diff_stat}F demonstrates that the $1/N$ scaling holds also for the diffusive motion over long time scales:
the diffusion coefficient, extracted from a fit to the statistics of an OU process, is inversely proportional to $N$. The same scaling with $N$ has been observed in continuous attractor networks with intrinsic neural stochasticity
\cite{burak_fundamental_2012}. Another important implication of the
$1/N$ scaling is that sufficiently large networks can reliably store a continuous variable in short-term memory (see \textit{Discussion}). 

\begin{figure}
\centering
\medskip
\includegraphics[scale=1.0]{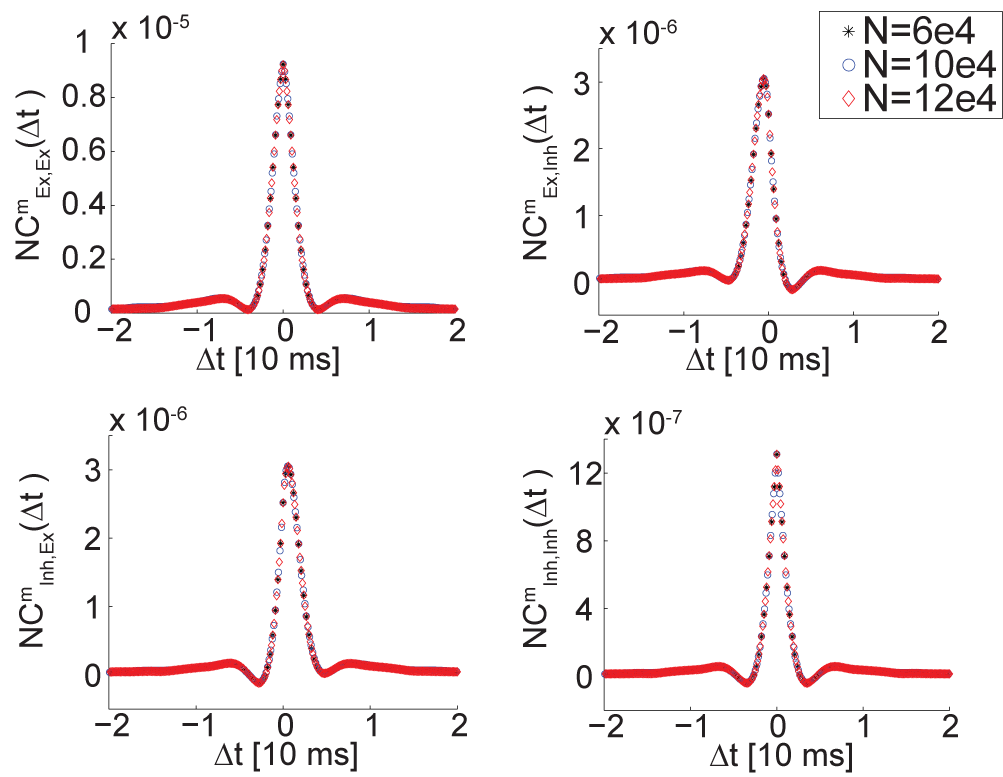}
\vspace{2mm}
\caption{\textbf{Cross correlations in a single balanced network}. Population averaged cross-correlations of neural activity in a single balanced network, $C^m(\Delta t)$ (Eq.~\ref{cm_single}),
shown as a function of the time lag $\Delta t$ (in units of $\tau = 10\,$ms). 
  The cross-correlation functions are multiplied by $N$, to demonstrate that $C_{ij}$ scale as $1/N$: with this choice of scaling, measurements from simulations with different values of N collapse on one curve. }
\label{fig:corr_vs_N}
\end{figure}

To understand this result in more detail, we start by considering
the time dependent correlation functions of $m_i$ in a \textit{single} balanced network:
\begin{equation}
C^m_{ij}(\Delta t) \equiv \frac{1}{N^2}  
\sum_{k,l=1}^N \left[
\langle \sigma_i^k(t+\Delta t) \sigma_j^l(t) \rangle_t - \langle \sigma_i^k(t+\Delta t) \rangle_t \langle \sigma_j^l(t) \rangle_t
\right]\,,
\label{cm_single}
\end{equation}
where 
 $i,j \in \{\text{Ex,Inh}\}$.
An analytical expression for these correlation functions is not available
(see \cite{renart2010asynchronous} for further discussion). 
Therefore, we 
measured them numerically in simulations of activity in the single balanced network architecture. 
These measurements indicate that the time-dependent correlation functions
decay over time scales of order $\tau$, and that they scale as $1/N$, Fig.~\ref{fig:corr_vs_N}.

Next, we show that the statistics of diffusion in the coupled system
can be expressed precisely in terms of the correlation functions of the single, uncoupled balanced networks 
(\textit{Methods}). 
Thus, the correlation structure of the chaotic noise in the single balanced
network determines the statistics of the slow diffusive motion along the approximate attractor in the 
coupled two-population network. 

Using the noise cross correlations measured in simulations of a single balanced network, 
it is possible to obtain a semi analytical approximation for $G$ in
the system of two coupled subnetworks, which does not involve
any fitting parameters.
The measurements of $G$ from simulations are in excellent agreement with this analytical prediction, Fig.~\ref{fig:diff_stat}E. 
The above analysis indicates that
the $1/N$ scaling of the diffusion coefficient (Fig.~\ref{fig:diff_stat}F) is a consequence of the decay with $N$ of cross correlations in activity of different neurons in the single balanced 
network. In this sense, for large $N$ the network behaves as a collection of neurons with independent random noise, 
although the 
source of this apparent noise is the chaotic activity generated by the recurrent connectivity. 

\subsection*{Spike correlation functions}

\begin{figure}
\centering
\medskip
\includegraphics[scale=1.0]{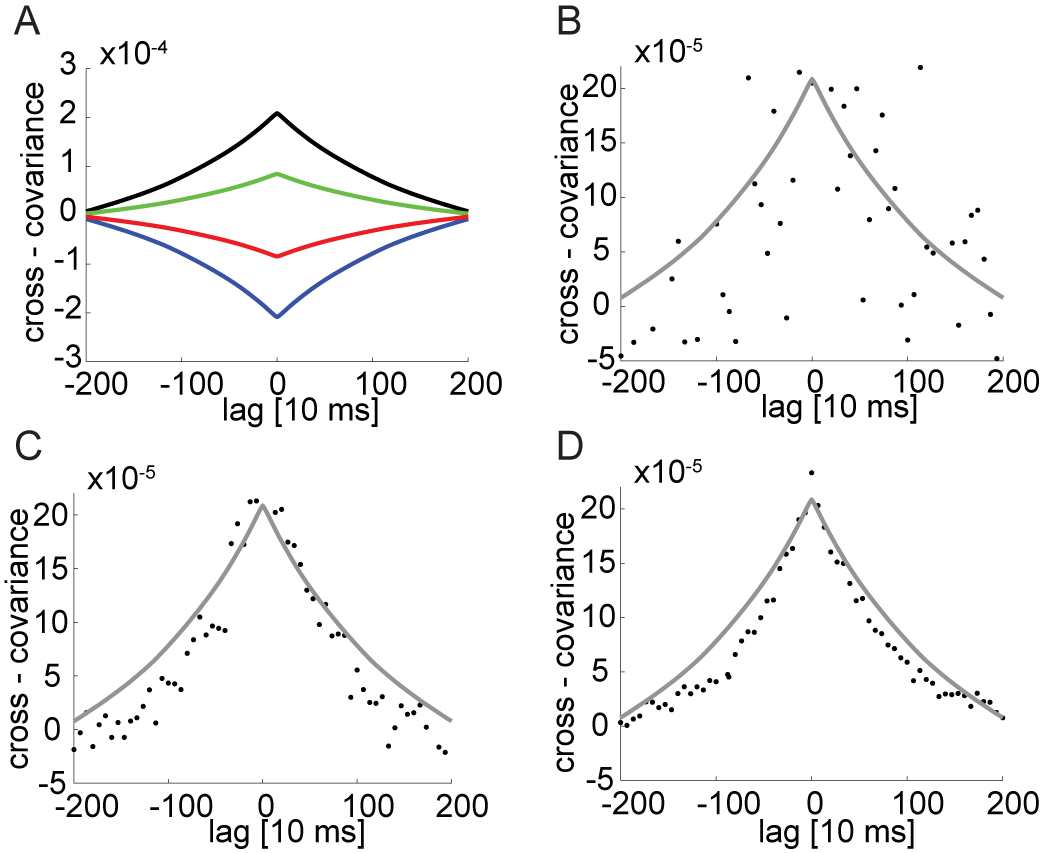}
\vspace{2mm}
\caption{\textbf{Cross correlations in the coupled subnetwork architecture}. \textbf{A} Auto-covariance of the mean activity of one of the excitatory populations (black). Cross covariance between the first excitatory population  and the first inhibitory population (green), the second inhibitory population (red) and the second excitatory population (blue). \textbf{B}-\textbf{D} Pairwise correlation of neuron activity in one of the excitatory populations. Correlations are measured in $n$ neurons,
selected randomly, and averaged over all pairs (\textit{Methods}), where $n = 10$ (B), 50 (C), and 100 (D).  
The simulated time is 15 minutes. An average over the entire population is shown in grey. In all panels $N=1.2\times 10^5$, $K=1000$, $\lambda^{-1}\simeq 2$ sec.}  
\label{fig:corr_all}
\end{figure}

The diffusion along the approximate attractor implies that the population
activities are correlated over long time scales, up to order $\lambda^{-1}$: Fig.~\ref{fig:corr_all}A shows
examples of the population correlation functions $C^m$, which differ dramatically from those
of the single balanced network, Fig.~\ref{fig:corr_vs_N} (note the different time scales in the two sets of figures). 

Spike trains generated by single neuron pairs are correlated over long time scales as well,
since all neurons in the network are coupled to the collective diffusion 
along the approximate attractor. However, a 
reliable 
observation of the slowly decaying correlation in a single pair might require an unrealistically long recording time. This difficulty can be overcome potentially by considering the simultaneous activity
of multiple neurons: for example, we find in our simulations of a network with $N =10^5 $ that for 15 minutes of simulated time, a simultaneous recording from $\sim$50 or more neurons from each population would be sufficient to reliably observe the slow temporal decay of the correlations, Fig.~\ref{fig:corr_all}~C,
whereas a simultaneous recording from ten neurons over 15 minutes may be insufficient. 
As demonstrated in Fig.~\ref{fig:corr_all}~(B-D) the noise falls as one over the number of measured neurons and as one over the total recording time. Hence, by extrapolating from the results in Fig.~\ref{fig:corr_all}(B-D), $\sim$12 hours of recording would be 
required to obtain a measurable correlation signal from a single pair of neurons.

\begin{figure}
\centering
\medskip
\includegraphics[scale=1.0]{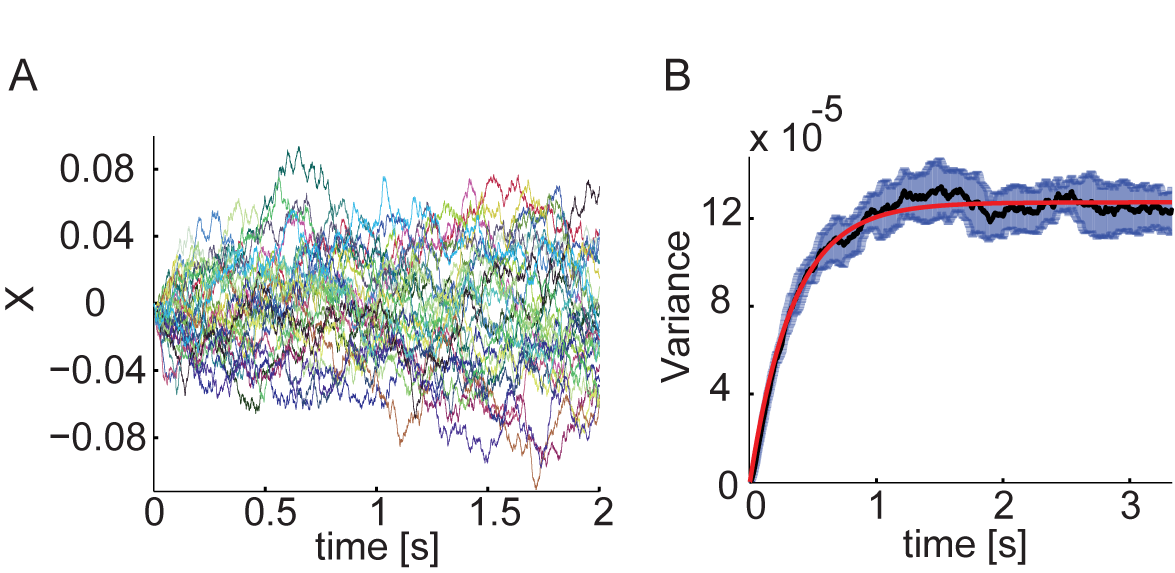} 
\vspace{2mm}
\caption{\textbf{Chaotic nature of the noise driving the diffusive motion}. \textbf{A} Projection $X$ of the mean activities on the approximate
attractor in 30 trials with the same update schedule and the same initial conditions, except for one neuron which was flipped in each population ($N=10^5$). \textbf{B} Variance over 1500 trials as a function of time, with 2$\sigma$ error bars (gray). Red: fit to the variance of an OU process ($D \simeq 3.4\cdot 10^{-6} 1/ 10\text{s}$ ).} 
\label{fig:chaos}
\end{figure}

\subsection*{Chaotic behavior}
Next, we 
briefly 
address the chaotic nature of the noise that drives diffusive motion. Figure \ref{fig:chaos}A shows results from multiple simulations in which the initial network state differed solely by a flip of one neuron in each population (out of $\sim 10^5$ neurons). All other parameters, including the asynchronous update schedule and the network weights were identical across runs.
The time dependence of the variance across different runs is similar to the variance over realizations of an OU process, Fig.~\ref{fig:chaos}B, with a similar diffusion coefficient as observed
in the fit for $G(X,\Delta t)$, Fig.~\ref{fig:diff_stat}(C,E).
Thus, the different initial conditions are equivalent to 
different realizations of dynamic noise that drives
diffusive motion along the approximate line attractor. 

\subsection*{Additional randomness in connectivity and inputs}
\textcolor{myblue}{In addition to the results described above, we investigate several scenarios in which we introduce additional randomness, either frozen or dynamic. First, we relax the assumption that connections in the two sub-networks precisely mirror each other. This assumption was made above for convenience: the precise identity of the synaptic connections
simplifies the numerical analysis since it ensures a precise symmetry of the dynamics around the hyperplane 
$X = 0$. S2~Fig demonstrates that the main conclusions of our analysis remain valid when the connectivity in each sub-network is drawn independently: dynamics are slow along the approximate line attractor, and the diffusion coefficient along the line scales as $1/N$ with a prefactor which is somewhat larger than the value observed in Fig.~\ref{fig:diff_stat}. }

\textcolor{myblue}{The main new feature that arises when the synaptic connections are drawn independently in the two sub-networks, 
is that the relaxation point of the dynamics along the approximate attractor deviates from the hyperplane $X = 0$. The characteristic magnitude of this deviation decays monotonically to zero with increase of the system
size $N$. We note that the mean field description of the dynamics
for finite $K \gg 1$ and in the limit $N\rightarrow \infty$, 
is identical to the mean field dynamics associated with the perfectly symmetric scenario.}

\textcolor{myblue}{Next, we consider a scenario in which the inhibitory connections between the two sub-networks are random, and follow  the same basic architecture as the connections within each sub-network. Therefore, instead of assuming weak all-to-all connections of order $\sqrt{K}/N$, we include random connections of order $1/\sqrt{K}$, with a probability $K/N$ for a synaptic connection. In addition, we relax the assumption of mirrored connections in the two sub-networks. In this case it is straightforward to show that the mean-field equations remain identical to those associated with the case of all-to-all connections, in the limit $N\rightarrow \infty,\; K \rightarrow \infty$. Therefore, a continuous set of balanced states can be achieved (\textit{Methods}).}

\textcolor{myblue}{When $K$ is large and finite, the mean-field equations are slightly different in the two scenarios (\textit{Methods}). The main outcome of this difference
is a shift of the unstable fixed points of the dynamics from the planes $m_1 = m_2 = 0$ and $m_3 = m_4 = 0$. Consequently, there is a certain degree of activity in both sub-networks even in the unstable fixed points of the dynamics. This is shown in S3~Fig. More significantly, S3~Fig demonstrates that in the case of random and sparse connections between the two sub-networks, the dynamics exhibit the same characteristics as in the case of all-to-all connectivity, and can be accurately approximated as an OU process over time scales longer than $\tau$. The coefficient of diffusion $D$ scales linearly with $1/N$, with a prefactor which is close to the one observed in S2~Fig.}

\textcolor{myblue}{Finally, we explore the effects of stochasticity in the input $E_0$ to the network (see Fig.\ref{fig:arch}B). S4~Fig demonstrates that even when the inputs include a large degree of temporal variability, and the noise injected to all the neurons in the network is highly correlated, the network exhibits slow dynamics along an approximate attractor, with statistics that are qualitatively similar to what we present above. }

\section*{Discussion}
In summary, we demonstrated that a simple balanced network can exhibit slow dynamics along a continuous line in the population mean activity space.  
In finite networks, the chaotic dynamics drive diffusive motion along the approximate line attractor.  
We calculated the diffusivity in the system, based on the correlation structure observed in a single balanced network, and showed that the diffusion coefficient along the approximate attractor is inversely proportional to the network size. This is similar to the effect of noise that arises from intrinsic neural or synaptic mechanisms \cite{burak_fundamental_2012}. 

The slow diffusive motion along a one-dimensional trajectory induces correlations within the populations, and in single neuron pairs, that persist up to a long time scale set by the decay time $\lambda^{-1}$. This property characterizes the dynamics of the system even when it is at the resting state
($X \simeq 0$), \textit{i.e.}, when the network is not engaged in a memory task. Hence, this observation generates a prediction for
spontaneous activity in
brain areas such as the prefrontal cortex, in which continuous attractor dynamics, based on mutual inhibition between two populations, have been postulated \cite{machens2005flexible}. 

A slowly decaying cross-correlation function characterizes the spikes produced by any pair of neurons in the network, but due to the high irregularity in the single neuron activity it may be necessary to average over multiple simultaneously 
recorded neuron pairs 
in order to obtain a clear measurement over a realistic time scale for a single experiment
(in Fig.~\ref{fig:corr_all}C, 50 neuron pairs and a $\sim$15 minute measurement). Furthermore, it is
important
to label the neurons based on their functional properties: averaged over all populations, the cross correlations seen in Fig.~\ref{fig:corr_all}A cancel. In brain areas involved in short term memory tasks, this labeling can potentially be achieved by first measuring the tuning 
curves of neurons as a function of the stored variable.

\subsubsection*{Linear \textit{vs}. nonlinear neural response in decision making circuits}

Several models of decision making circuits in the prefrontal cortex were based on
the simple neural architecture of Fig.~\ref{fig:arch}A 
\cite{wang2002probabilistic}. 
This network architecture can precisely generate a continuous attractor if the activity of single units 
is a linear function of their input.
While the linear dynamics provide a simple
intuition for the principles underlying continuous attractor dynamics in recurrent
neural networks \cite{cannon1983proposed}, 
it is more difficult to obtain a continuum of steady states using the above 
network architecture when single neural responses are nonlinear. Therefore, specifically tuned forms of nonlinearity
\cite{machens2005flexible,Doiron2012variability}, or 
more elaborate network
architectures -- still based on mutual inhibition between two or more neural populations
have been proposed \cite{machens2005flexible,deco2006decision}. 
In this context the linear input-output relationship, characterizing
the single balanced network of Ref.~\cite{van1996chaos}, is a useful computational feature that
facilitates the construction of a continuous attractor network based on the simple architecture of Fig.~\ref{fig:arch}A.
However, the main motivation for considering the balanced state in this work lies in its ability
to account for the irregular spiking of single neurons in cortical circuits.

\subsubsection*{Time scales of persistence and retention of information}
Continuous attractor networks are an important model for maintenance of short-term memory in the brain.
The memory is 
represented by the position along the attractor, and therefore the stochastic motion
along the attractor determines the fidelity of memory retention. 
{\color{myblue} 
Since the dynamics of our proposed network are well characterized as an OU process over time scales longer than $\tau$ and over a large
range of positions along the approximate attractor, it is straightforward to assess how the 
position along the approximate attractor evolves in time. 
All aspects of the trajectory can be easily inferred based on the 
initial state along the approximate attractor, the time interval, and the two parameters which
characterize the OU process: $\lambda$ and $D$. Similar considerations can be applied
also for noisy continuous attractors in which the stochasticity arises from mechanisms other than the chaotic dynamics studied here \cite{burak_fundamental_2012, kilpatrick2013optimizing}.
We next discuss how the 
decay time and the diffusivity depend on the parameters of our model.}

The decay time $\lambda^{-1}$ can be calculated exactly in the limit of $N\rightarrow \infty,\; K \gg 1$
(Fig.~\ref{fig:finite_K}). {\color{myblue} It is interesting to note that there are competing influences of $K$ on the tuning of the attractor: with increase of $K$, $\lambda^{-1}$ becomes more sensitive to $\tilde{J}$. However, when $K$ is reduced, the nullclines (Fig.~\ref{fig:finite_K}) become less linear, causing deviations from the ideal behavior
far from the symmetry point. We showed that for $K \sim 10^3$ and $\tau = 10\,$ms it is possible to achieve persistence over several seconds, while the decay time and diffusion coefficient are approximately constant along a wide range of positions. This requires to tune $\tilde{J}$ to a relative precision of order 0.1$\%$.

Since all times scale linearly with the intrinsic time constant, longer persistence times (or a weaker tuning requirement) can be achieved if the intrinsic time constants of individual units is longer that the value of 10\,ms assumed in our examples. Intrinsic neural persistence or slow synapses could potentially contribute to this goal under more realistic biophysical descriptions of the neural dynamics.
}
 
Finally, we note that the requirement for precise tuning of the connectivity is a characteristic feature of all continuous attractor models. Several works have proposed ways to achieve tuning through 
plasticity mechanisms
\cite{seung1997learning,macneil2011fine}, or ways to stabilize
the dynamics by additional mechanisms such as synaptic adaptation \cite{mongillo2008synaptic,hansel_short-term_2013,itskov2011short} or negative derivative feedback 
\cite{lim_balanced_2013,lim_balanced_2014}, in order to increase the persistence time. 

\textbf{Diffusive motion}\ \ The diffusive motion along the approximate attractor, which is the main focus of this work,
poses an additional limitation on the persistence of short term memory. 
{\color{myblue} While appropriate readout mechanisms may be able to take into account the systematic drift caused by the decay towards to the symmetry
point, random diffusion inherently degrades the information stored
in the position along the attractor.

With $10^5$ neurons per population, random diffusion over an interval of
one second causes a deflection in $X$ with a standard deviation of $\sim 10^{-2}$. This quantity should be compared with the possible range of $X$, which is approximately [-0.2,0.2] in our parametrization of the 
position along the approximate attractor (we verified that the dynamics are accurately approximated as an OU process over the range [-0.1,0.1]). Therefore, with the 
tuning chosen in Fig.~\ref{fig:finite_dyn}C, where $\lambda^{-1} \sim10\,$s, and with $N = 10^5$, the limiting factor for discrimination between nearby stimuli after a delay period of order 1\,s is the diffusive dynamics along the approximate attractor.

Random diffusion in continuous attractor networks of Poisson neurons is also often very significant \cite{burak_fundamental_2012,kilpatrick2013optimizing}. The diffusivity can be suppressed by increasing the number of neurons, increasing the intrinsic time constant of individual neurons and synapses, or by assuming that the firing of individual neurons is sub-Poisson
\cite{burak2009accurate,burak_fundamental_2012}. Our proposed model for a line of persistent balanced states similarly predicts a significant degree of random diffusion, highlighting the need to better understand
how noise influences the retention of continuous parameter memory in cortical circuits. 

There are several ways in which the random, diffusive component of the motion can potentially be reduced: First, by increasing the number of neurons. We presented results for networks containing (altogether) up to 
$6\times 10^5$ neurons, and it is straightforward to extrapolate our estimates for $D$ to larger networks based on the $1/N$ dependence of the diffusion coefficient. Second, the diffusion coefficient is expected to decrease significantly if slow synapses participate in the dynamics \cite{burak_fundamental_2012}, or if the intrinsic time constant of the neurons is increased. Third, additional mechanisms such as synaptic adaptation \cite{itskov2011short,hansel_short-term_2013} or derivative feedback \cite{lim_balanced_2013} may perhaps contribute to a reduction in the diffusivity. Finally, an intriguing possibility is that highly structured and tuned connectivity can yield improved robustness to noise in a balanced state, as hinted by recent results on predictive coding in spiking neural networks \cite{boerlin2013predictive}. }

\section*{Methods}

\subsection*{Model}

In our model, two balanced neural subnetworks inhibit each other reciprocally: 
the inhibitory population in each subnetwork projects to the excitatory population of the other subnetwork, Fig.~\ref{fig:arch}B. 
 
As in Refs. \cite{van1996chaos,van_vreeswijk_chaotic_1998},
the neurons are binary and are updated asynchronously, at update times that follow Poisson statistics.
The mean time interval between updates is $\tau_E$ ($\tau_I$) for neurons in the excitatory (inhibitory) populations.
In each update of a neuron $k$ from population $i$, the new state of the neuron $\sigma_i^k$ is determined based on the total weighted input 
to the neuron,
\begin{equation}
  \sigma_i^k=\Theta(u_i^k) \,,
\end{equation}
where $\Theta$ is the Heaviside step function, and $u_i^k$ is the total input to the unit at that time,
\begin{equation}
  u_i^k=\sum_{l=1}^{4} \left[\sum_{j=1}^{N_l} J_{kl}^{ij} \sigma_l^j(t)+\sqrt{K} E^l_0\right]-T_k\,.
\end{equation}
Here, $T_k$ is the threshold and $E_0$ is an external input. We chose the external input 
to be zero for the inhibitory populations and to be positive (and constant) for the excitatory populations.
Connections within each network are random with a connection probability $K/N$,  where $1\ll K \ll N$. Here $N$ is the population size (chosen to be equal in all populations for simplicity) and $K$ is the average number of inputs a neuron gets from each population. Connection strengths are:  $ J_{EE}/\sqrt{K},\:   J_{IE}/\sqrt{K}, \: J_{EI}/\sqrt{K}$ and $ J_{II}/\sqrt{K} $ according to the identity of the participating neurons. Without loss of generality, we 
have chosen $J_{EE}=J_{IE}=1$ and defined $J_{EI}\equiv -J_E, J_{II}\equiv -J_I$. 
Mutual inhibition is generated either by weak all-to-all connections (in all figures except for S3~Fig), or by
strong random and sparse connections (S3~Fig). In the former scenario, synapses of strength 
$-\tilde{J}\sqrt{K}/N$ connect each inhibitory neuron to all excitatory neurons in the excitatory population of the other subnetwork. In the latter scenario
(S3~Fig),
connections from each inhibitory population to the excitatory population of the other subnetwork are chosen randomly with a connection probability 
$K/N$, and with strength $-\tilde{J}/\sqrt{K}$.
An excitatory feed-forward input 
$\sqrt{K} E_0$ is fed into both excitatory populations.
We denote by $u_i$ the mean of $u_i^k$ over all the neurons $k$ within the population $i$ and over realizations of the connectivity:
\begin{equation}
\begin{array}{rcl}
u_1 &=& \sqrt{K}(m_1-J_E m_2-\tilde{J} m_4 +E_0)-T_1 \,,\\
u_2 &=& \sqrt{K}(m_1-J_I m_2)-T_2 \,,\\
u_3 &=& \sqrt{K}(m_3-J_E m_4-\tilde{J} m_2 +E_0)-T_3 \,,\\
u_4 &=& \sqrt{K}(m_3-J_I m_4)-T_4 \,.
\end{array}
\label{mean_input}
\end{equation}
Here $m_i$ are the population averaged activities. The variance of $u_i^k$ over all the neurons $k$ within the population $i$ and over realizations of the connectivity is given (to leading order in $K/N$) by:
\begin{equation}
\begin{array}{rcl}
\alpha_1 &=& m_1+J_E^2 m_2 \,,\\
\alpha_2 &=& m_1+J_I^2 m_2 \,,\\
\alpha_3 &=& m_3+J_E^2 m_4 \,,\\
\alpha_4 &=& m_3+J_I^2 m_4\,.
\end{array}
\label{alpha_all_to_all} 
\end{equation}
These expressions are obtained in similarity to the derivation of the variances in Ref. \cite{van_vreeswijk_chaotic_1998}. Note that the all-to-all inhibitory connections between the subnetworks contribute only terms of higher order in $K/N$. {\color{myblue}In the scenario where the connections between subnetworks are randomly drawn (S3~Fig), the variance of the input to the excitatory neurons includes an additional term, due to the variability of inhibitory synapses from the opposing sub-network. In this scenario
\begin{equation}
\begin{array}{rcl}
\alpha_1 &=& m_1+J_E^2 m_2+\tilde{J}^2 m_4 \,,\\
\alpha_2 &=& m_1+J_I^2 m_2 \,,\\
\alpha_3 &=& m_3+J_E^2 m_4+\tilde{J}^2 m_2 \,,\\
\alpha_4 &=& m_3+J_I^2 m_4\,.
\end{array}
\label{alpha_random}
\end{equation}
\noindent The mean field equations written below are valid both for all-to-all and for random connections between sub-networks, with the appropriate choice of $\alpha_i$.}

\subsubsection*{Line of balanced states in the limit $N \gg K \gg 1$}

To check whether there exist parameters for which the system has a continuum of balanced states, it is convenient to write the 
steady state of equation (\ref{full_diff_sys}) as follows:
\begin{equation}
\begin{array}{rcl}
	m_1-J_E m_2-\tilde{J} m_4 +E_0 &=& \frac{1}{\sqrt{K}}\left[T1-\sqrt{\alpha_1}H^{-1}(m_1)\right] \,,\\
	m_1-J_I m_2 &=& \frac{1}{\sqrt{K}}\left[T2-\sqrt{\alpha_2}H^{-1}(m_2)\right] \,,\\
	m_3-J_E m_4-\tilde{J} m_2 +E_0 &=& \frac{1}{\sqrt{K}}\left[T3-\sqrt{\alpha_3}H^{-1}(m_3)\right] \,,\\
	m_3-J_I m_4 &=& \frac{1}{\sqrt{K}}\left[T4-\sqrt{\alpha_4}H^{-1}(m_4)\right] \,.
\end{array}
\label{full_sys}
\end{equation}
Taking the limit $K\rightarrow \infty$ while requiring that none of the populations is fully on or off produces a linear system 
of equations for the mean activities, Eq.~\ref{full_sys_lin}. 
When $\tilde{J}=J_E-J_I$
the system of linear equations 
is singular. In this case the steady state equations
admit a continuum of solutions which comprise a continuum of stable balanced states. 
A possible parametrization of the line of balanced states is given by:
\begin{eqnarray}
m_1 & = & x \,, \nonumber \\
m_2 & = & x/J_I \,, \nonumber \\
m_3 & = & -x+J_I E_0/(J_E-J_I)\,, \nonumber \\
m_4 & = & -x/J_I+ E_0/(J_E-J_I) \,.
\label{fixed_line}
\end{eqnarray}
The conditions $J_E-J_I >0$, $J_I>1$, and $0<J_I E_0/(J_E-J_I) <1$ ensure that for $0<x<J_I E_0/(J_E-J_I)$ the mean activities are positive and none of them is equal to $0$ or $1$. 

In Fig.~\ref{fig:finite_K}(E-F), we artificially add to the dynamics  (Eq. \ref{full_diff_sys}) white Gaussian noise as follows:
\begin{equation}
\tau_i \dot{m}_i=-m_i+H(u_i/\sqrt{\alpha_i})+\xi_i
\label{noise_dyn}
\end{equation}
where $\left\langle\xi_i(t)\right\rangle = 0$ and
\begin{equation}
    \left\langle\xi_i(t)\xi_j(t') \right\rangle= \sigma^2 \delta_{ij} \delta(t-t')\,.
\label{white_noise}
\end{equation}
Note that this was done only in Figs.~\ref{fig:finite_K}(E-F), to illustrate the existence of slow dynamics along a line in the case of infinite
$N$. There is no injected noise elsewhere, and in particular there is no injected noise in our simulations
of finite $N$ networks. 


\subsection*{Simulations and statistics of the diffusive dynamics}

Our results for networks with finite $N$ are based on large scale numerical simulations. In each simulation the connections were chosen randomly as described in the text, and an asynchronous update schedule was generated by a Poisson process. Parameter values 
are specified in the legend of Fig.~\ref{fig:finite_K} in the main text.  Averaged population activities were calculated online. The projection along the approximate attractor was defined at each time point as 
\begin{equation}
X(t) \equiv \boldsymbol{v}_0^T\cdot \left[\boldsymbol{m}(t) - \boldsymbol{m}_0\right]\,,
\label{proj_def}
\end{equation}
where $\boldsymbol{m}(t)$ is the measured 4 dimensional averaged population activity, $\boldsymbol{m}_0$ is the vector of mean population activities at the symmetric fixed point, and $\boldsymbol{v}_0$ is the left eigenvector of the linearized dynamics with an eigenvalue close to zero. We chose the following normalization for the corresponding right eigenvector (see eq.~\ref{fixed_line}):
\begin{equation}
    \left( \begin{array}{c} 1\\ 1/J_I \\ -1\\ -1/J_I \end{array} \right) \,,
\end{equation}
and the normalization of $\boldsymbol{v}_0$ was chosen such that the dot product of the left eigenvector and the right eigenvector equals unity.

Measurements of $G(X,\Delta t)$ (Eq.~\ref{diffusion1}) were done in the following way: for each value of $X$ we found all the time points for which $\left | X(t_i)-X \right |<\delta$, using a small $\delta \simeq 10^{-3}$. Then, for each such $t_i$ we calculated $\left[X(t_i+\Delta t)-X(t_i)\right]^2$, and averaged all these values to get $G(X,\Delta t)$. Subsequently, we averaged over multiple simulations with different quenched noise and update schedules. In the manuscript we present results for $G(0,\Delta t)$, but
in similarity to $F(X,\Delta t)/X$, $G(X,\Delta t)$ was fairly uniform along
the approximate attractor. 
A similar calculation was performed to measure the drift $F(X,\Delta t$). 
In Figs.~\ref{fig:diff_stat}(C-F), results are based on $(1-2) \times 10^3$ simulations with random initial conditions, each spanning a simulated time 
of about 10 seconds.
 In simulations of the finite $N$ network, we estimated $\lambda$ from measurements of $F(X,\Delta t)$ near the symmetric 
fixed point, and tuned $\tilde{J}$ to obtain $\lambda^{-1} \gg \tau$. 
In Fig.~\ref{fig:diff_stat}, $\lambda^{-1} \simeq 2$\, seconds.

\subsubsection*{Measurement of cross covariance functions}
The cross covariance functions shown in Fig.~\ref{fig:corr_all} were calculated in the following way: 
first, we measured the activity of $n$ neurons at $M$ equally spaced time points, with a time difference $\Delta t = 66\;\text{ms}$.
Then, for each pair of neurons $i,j$ of populations $k,l$ respectively, we calculated the unbiased estimate of the cross covariance:
\begin{equation}
C_{l,k}^{i,j}(t_m)= \frac{1}{M-|m|}\sum_{a=0}^{M-|m|-1} \sigma_k^i(t_{a+m}) \sigma_l^j(t_a)-\left[\frac{1}{M}\sum_{a=0}^{M-1} \sigma_k^i(t_a) \right] \left [\frac{1}{M}\sum_{a=0}^{M-1} \sigma_l^j(t_a) \right]\, ,
\end{equation}
Here $t_a=a \Delta t$.
Now, we averaged over all the measured pairs:
\begin{equation}
C_{l,k}(t_m)=\frac{1}{0.5n(n-1)} \sum_{i\neq j} C_{l,k}^{i,j}(t_m)\,.
\label{pair_corr_calc}
\end{equation}
For the calculation of the entire population averaged cross covariance we used the measured mean activities:
\begin{equation}
C_{l,k}(t_m)= \frac{1}{M-|m|}\sum_{a=0}^{M-|m|-1} m_k(t_{a+m}) m_l(t_a)-\left[\frac{1}{M}\sum_{a=0}^{M-1} m_k(t_a) \right] \left[\frac{1}{M}\sum_{a=0}^{M-1} m_l(t_a) \right]\, .
\label{m_corr_calc}
\end{equation}
where $m_l(t)=1/N\sum_{i=1}^N \sigma_l^i(t)$.
Note that the sum in Eq.~\ref{m_corr_calc} includes the auto-covariances, while the expression in Eq.~\ref{pair_corr_calc} does not. However, the contribution of the auto-covariances is negligible in the entire population average, since its contribution, relative to the contribution of cross-covariances scales as $1/N$. 

\subsection*{Diffusion over short time scales}

To analytically evaluate $G(X,\Delta t)$ (Eq.~\ref{diffusion1}) over short time scales, 
we start by writing the change in the state of the $k$-th neuron in population $i$ in a short
time interval $\Delta t$ as:
\begin{equation}
\sigma^k_i(t+\Delta t)-\sigma^k_i(t)=c^k_i(t)\left[\Theta^k_i(t)-\sigma^k_i(t) \right]\,,
\end{equation}
where $\Theta^k_i(t)$ is the outcome of an update if it occurs, and
 $c^k_i(t)$ is a random variable equal to $1$ if the $i$-th neuron was updated between $t$ and $t+\Delta t$ and to $0$ otherwise. The updates occur each $\tau$ ms on average, so that
\begin{equation}
\left\langle c^k_i(t) \right\rangle_t =\left\langle c^k_i(t)^2 \right\rangle_t =\frac{\Delta t}{\tau_k}\,,
\label{c_ii}
\end{equation}
whereas for $i\neq j$ and/or $k \neq l$,
\begin{equation}
\left\langle c^k_i(t)c^l_j(t) \right\rangle_t = \frac{\left(\Delta t\right)^2}{\tau_k \tau_l}\,.
\label{c_ij}
\end{equation}
Now the mean squared displacement, $G(X,\Delta t)$, can be written as:
%
\begin{equation}
\left\langle\left[X(t+\Delta t)-X(t)\right]^2\right\rangle =\frac{1}{N^2}\sum_{i,j=1}^4 \sum_{k,l=1}^N v_i^0 v_j^0 
\left\langle\left[\sigma^k_i(t+\Delta t)-\sigma^k_i(t)\right]\left[\sigma^l_j(t+\Delta t)-\sigma^l_j(t)\right] \right\rangle\, ,
\end{equation}
%
where $v_i^0$ is the i'th component of the left eigenvector of the Jacobian with eigenvalue close to zero.
From Eqs.~\ref{c_ii}-\ref{c_ij} we see that for $\Delta t \ll \tau$ the contribution of elements with $i=j,\, k=l$ dominates the sum. To leading order in $\Delta t$ we have
\begin{eqnarray}
\left\langle\left[X(t+\Delta t)-X(t)\right]^2\right\rangle & \simeq & \frac{1}{N^2}\sum_{i=1}^4 \sum_{k=1}^N \left( v_i^0 \right)^2
\\ 
& \times & \left\langle\left[\sigma^k_i(t+\Delta t)-\sigma^k_i(t)\right]^2 \right\rangle\,.
\nonumber
\end{eqnarray}
Defining
\begin{equation}
q_i(\Delta t)=\frac{1}{N}\sum_{k=1}^N \left\langle \sigma^k_i(t+\Delta t)\sigma^k_i(t) \right\rangle\,,
\end{equation}
we obtain:
\begin{equation}
G(X,\Delta t)\simeq \frac{2 \Delta t}{N}\sum_{j=1}^4 (v^0_j)^2 \left.\left(-\frac{\partial q_j(t)}{\partial t}\right) 
\right|_{t\rightarrow 0} \,.
\label{analytical_short}
\end{equation}
%
\subsection*{Diffusion over arbitrary time scales}

To derive an expression for the diffusive dynamics over arbitrary time scales, we 
start by representing 
the stochastic linearized dynamics of a single balanced network, near the symmetric fixed point, as 
a two dimensional stochastic process:
\begin{equation}
\begin{array}{c}
\dot{\boldsymbol{\delta m}}=B_1 \boldsymbol{\delta m}+B_2 \boldsymbol{\delta E}+\boldsymbol{\xi}\,,
\end{array}
\label{l_response}
\end{equation}
where $\boldsymbol{\delta m}$ is the deviation of the mean activities from the fixed point and $\boldsymbol{\delta E}$ is the deviation of the input from the constant input $E_0$.
Here $B_1$ is a $2 \times 2$ matrix representing the response to perturbations in $\boldsymbol{m}$, and $B_2$ is a $2 \times 2$ matrix representing the response
to perturbations in the feedforward input. Both are obtained analytically from a linearization of the mean
field dynamics. Finally, $\boldsymbol{\xi}$ is a random process with vanishing mean, whose covariance functions $C_\xi(\Delta t)$ 
are stationary and are yet unspecified:
\begin{equation}
C_{\xi,ij}(t-t') \equiv \left\langle \xi_i (t) \xi_j(t') \right \rangle \,.
\label{corr_c_xi_def}
\end{equation}
Using  
Eq.~\ref{l_response},
it is straightforward to relate $C_{\xi}(t)$ to the covariance of the activities (while assuming constant feedforward input, $\delta \bar{E} = 0$):
%
\begin{eqnarray}
C_{\xi}(t) & = & -\frac{{\rm d} ^2}{{\rm d}t^2}C^m(t) 
+ \frac{{\rm d}}{{\rm d}t}\left[C^m(t)B_1^T-B_1C^m(t)\right] \nonumber \\
& + & B_1C^m(t)B_1^T\,.
\label{corr_eq}
\end{eqnarray}
Using the measurements of $C^m$ from simulations, we can thus obtain $C^\xi$ numerically, using the above equation. In similarity to $C^m$, 
$C^\xi$ decays to zero over a time scale of order $\tau$.
Altogether, Eq.~\ref{l_response} describes the stochastic dynamics of a single balanced network close to the balanced state, in response to 
small fluctuations $\boldsymbol{\delta{E}}$ in the feedforward inputs. 
In the two-subnetwork architecture, each subnetwork is coupled only to the mean activity of the other
subnetwork, because of the all-to-all connectivity. More specifically, the mean activity of each
subnetwork linearly modulates the external input to the excitatory population of the other subnetwork. Therefore, we can approximate 
the state of the 4-population network as a stochastic process with the following 
dynamics:
\begin{equation}
 \dot{\boldsymbol{\delta m}}=A\boldsymbol{\delta m}+\boldsymbol{\xi}\,,
 \label{full_diff_sys1}
\end{equation}
where $\boldsymbol{\delta{m}}$ is now a 4 dimensional vector, whose first (last) two entries represent the state
of the first (second) subnetwork, 
and
$A$ is the Jacobian of the full 4 dimensional 
dynamics around the fixed point (Eq.~\ref{Jac}), related in a simple
manner also to the matrices $B_{1,2}$ defined above. The correlation matrix of the 4 dimensional 
noise vector $\boldsymbol{\xi}$ 
is given by
\begin{equation}
 \tilde{C}_{\xi}=\left( \begin{array}{cc}
C_{\xi} & 0  \\
0 & C_{\xi}  \end{array} \right)\,,
\label{corr_mat}
\end{equation}
where $C_{\xi}$ is the $2 \times 2$ noise correlation matrix (\ref{corr_c_xi_def}) evalulated for a single balanced network receiving fixed excitatory input, equal to the mean input to each subnetwork at the symmetric fixed point.
Finally, we use this description of the dynamics to predict the statistics of diffusion along the line. We multiply equation \ref{full_diff_sys1} from the left by $\boldsymbol{v}_0$, the eigenvector with the eigenvalue close to zero,
which we denote by $\lambda$ (note below that $\lambda < 0$):
\begin{equation}
\dot{X}=\lambda X+\boldsymbol{v}_0^{\rm T}\cdot \boldsymbol{\xi}\,.
\end{equation}
Here, $X=\boldsymbol{v}_0^{\rm T}\cdot  \boldsymbol{\delta m}$. Thus, we obtain using the Wiener - Khintchine theorem
the time dependent correlation function of $X$,
\begin{equation}
C_X(t)=-\frac{1}{2\lambda}\int_{-\infty}^{\infty}  e^{\lambda |t'|} \boldsymbol{v}_0^{\rm T} \tilde{C}_{\xi}(t-t') \boldsymbol{v}_0 {\rm d}t'\,.
\end{equation}
Finally, the diffusion over an arbitrary time interval $\Delta t$ is given by:
\begin{equation}
\left\langle \left[X(t + \Delta t)-X(t)\right]^2 \right\rangle=2\left[C_X(0)-C_X(\Delta t)\right]\,.
\label{semi_analytical}
\end{equation}

\subsection*{Proof: $\partial m_1/\partial m_3=-1 \leftrightarrow$ vanishing eigenvalue }
Here we show that when $\partial m_1/\partial m_3=-1$ at the symmetric point ($m_1=m_3, \; m_2=m_4$),  the Jacobian matrix has a vanishing eigenvalue, leading to slow dynamics near the fixed point.
 We denote: 
    \begin{equation}
     f_{i,j}\equiv \frac{\partial H\left(-u_i/\sqrt{\alpha_i}\right)}{\partial m_j} \,.
    \end{equation}
    In terms of these quantities, the Jacobian matrix can be written as
    \begin{equation}
    A \equiv
    \begin{pmatrix}
     f_{1,1}-1 & f_{1,2} & 0 & f_{1,4} \\
     f_{2,1}/\tau & (f_{2,2}-1)/\tau & 0 & 0 \\
     0 & f_{3,2} & f_{3,3}-1 & f_{3,4} \\
     0 & 0 & f_{4,3}/\tau & (f_{4,4}-1)/\tau \\
    \end{pmatrix}\,.
    \label{Jac}
   \end{equation}
    At the symmetric fixed point $f_{1,1}=f_{3,3}$, $f_{2,2}=f_{4,4}$, $f_{1,2}=f_{3,4}$, $f_{1,4}=f_{3,2}$, and $f_{2,1}=f_{4,3}$.
    The Jacobian's eigenvalues at that point are then:
 \begin{equation}
     \begin{array}{c}
     \lambda_{\pm}^{\pm}=\frac{1}{2} \left[(f_{1,1}-1)+\frac{f_{2,2}-1}{\tau}\right] 
      \\ 
      \pm  \frac{1}{2} \sqrt{\left((f_{1,1}-1)-\frac{f_{2,2}-1}{\tau}\right)^2+\frac{4}{\tau} f_{1,2} f_{2,1} \pm \frac{4}{\tau} f_{1,4} f_{2,1}}\,.
      \end{array}
      \label{eig_val}
 \end{equation}
    
    Next, we approximate the derivative $\partial m_1/ \partial m_3$ at the symmetric point. We use a first order Taylor expansion of the mean field equations to get:
     \begin{eqnarray}
        \delta m_1 & = & f_{1,1} \delta m_1 + f_{1,2} \delta m_2 +f_{1,4} \delta m_4 \,, \nonumber \\
        \delta m_2 & = & f_{2,1} \delta m_1 + f_{2,2} \delta m_2  \,,\nonumber \\
        \delta m_3 & = & f_{1,4} \delta m_2 + f_{1,1} \delta m_3 +f_{1,2} \delta m_4 \,,\nonumber \\
        \delta m_4 & = & f_{2,2} \delta m_4 + f_{2,1} \delta m_3 \,.
    \end{eqnarray}
       Here, $\delta m_i$ are the small deviations from the symmetric fixed point. Using these equations we can write $\delta m_2$ and $\delta m_4$ as functions of 
       $\delta m_1$ and $\delta m_3$:
       \begin{equation}
          \delta m_2 = \frac{f_{2,1}}{1-f_{2,2}}\delta m_1 \ \ \ ; \ \ \  \delta m_4 = \frac{f_{2,1}}{1-f_{2,2}}\delta m_3
         \,.
         \end{equation}
      Plugging these expressions into the equation for $\delta m_1$ yields an expression for $\delta m_1$ as a function of $\delta m_3$. The derivative is:
    \begin{equation}
     \frac{\partial m_1}{\partial m_3}= \frac{f_{1,4} f_{2,1}}{f_{1,1} f_{2,2}-f_{1,2} f_{2,1}+1-(f_{1,1}+f_{2,2})}\,.
    \end{equation}
    Note that $\partial m_1/\partial m_3=\partial m_2/\partial m_4$.
Equating this derivative to $-1$ (noting that in this case, $\partial m_1/\partial m_3=\partial m_3/\partial m_1$) yields:
\begin{equation}
      f_{1,4}= \frac{f_{1,2} f_{2,1}-1+f_{1,1}+f_{2,2}-f_{1,1} f_{2,2}}{f_{2,1}}\,.
    \end{equation}
    By inserting $f_{1,4}$ into \ref{eig_val} we get $\lambda_-^- =0$. 



\section*{Acknowledgments}
We thank Sophie Deneve for the helpful discussion which has motivated this work, and Haim Sompolinsky for comments on the manuscript.  

\nolinenumbers

%
%
%

\pagebreak
\section*{Supporting Figures}
\renewcommand{\thefigure}{S\arabic{figure}}
\setcounter{figure}{0}

\pagebreak


\begin{figure}
\centering
\medskip
\includegraphics[scale=1.0]{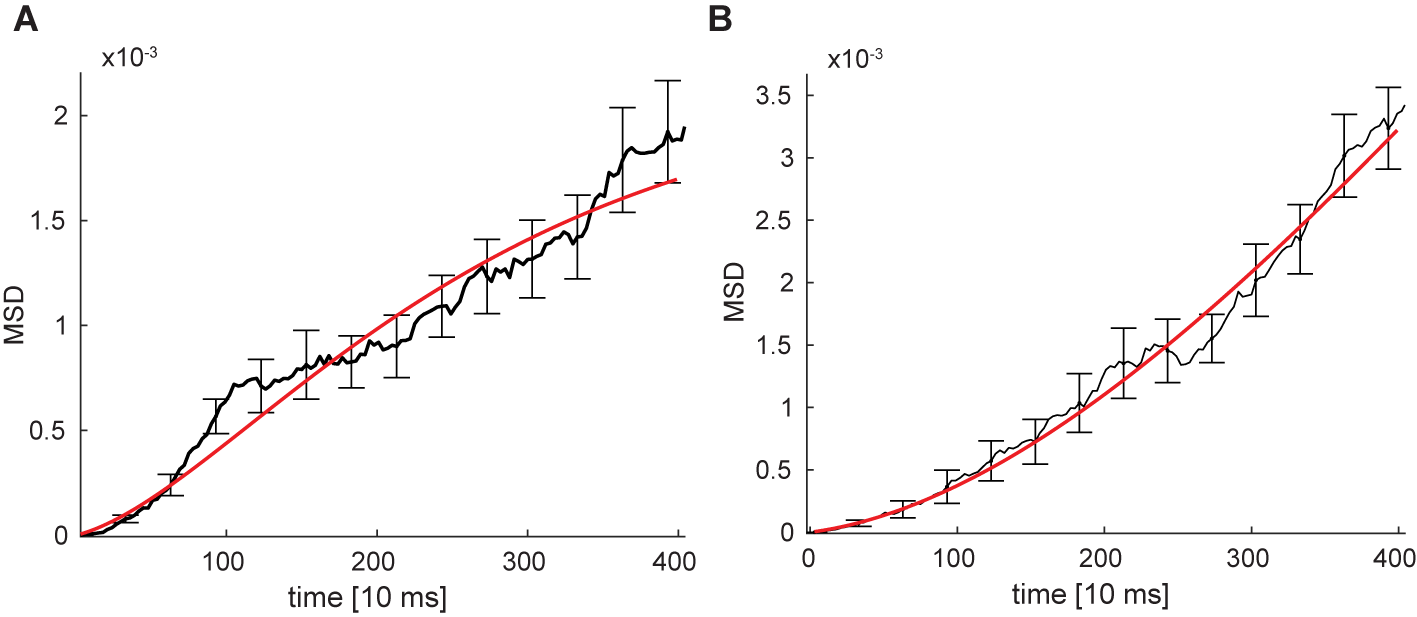}
\vspace{2mm}
\caption{\textbf{ Mean squared displacement of location along the approximate attractor.}  (Same dataset as in Fig.4C in the main text.) \textbf{A}-\textbf{B} The mean squared displacement (MSD) of the location along the line for initial location $X(0)=0.05$ (\textbf{A}) and $X(0)=0.08$ (\textbf{B}) as a function of time. Error bars represent the standard deviation of the mean (black). Red: fit to an OU process. OU parameters from fit: $D=1.85\times 10^{-6}\, (\text{10 ms})^{-1}$ , $\lambda= 10^{-3} \,(\text{10 ms})^{-1}$ in both panels.
}
\label{fig:MSD_two_points}
\end{figure}

\begin{figure}
\centering
\medskip
\includegraphics[scale=1.0]{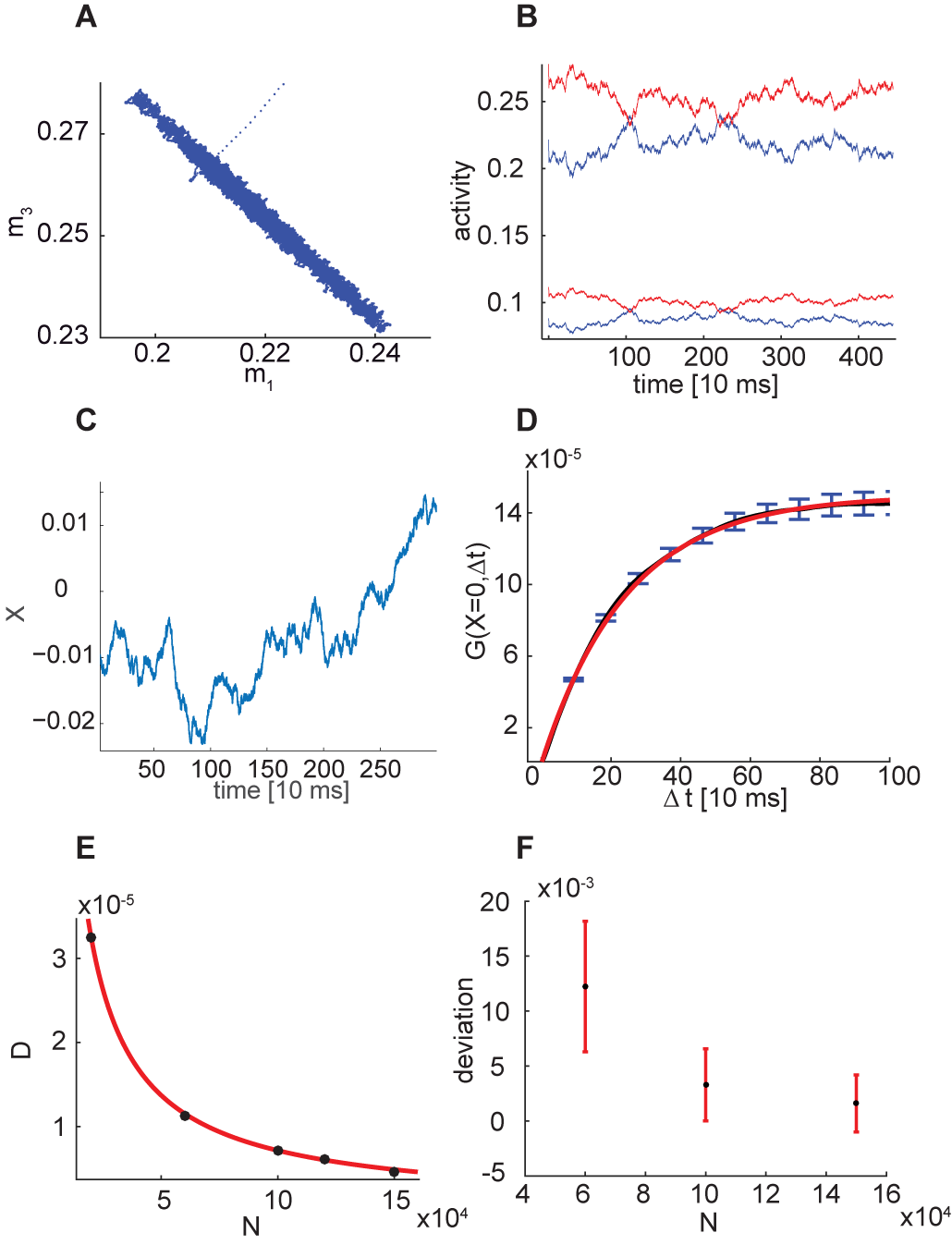}
\vspace{2mm}
\caption{\textbf{ Non-mirrored connectivity.} Results for a network in which the internal connectivity in each sub-network is drawn independently. \textbf{A} Population averaged activity projected onto the $m_1 - m_3$ plane. \textbf{B} Mean activities of the four populations: blue for one sub-network and red for the other. The higher activities are those of the excitatory populations ($m_1$ and $m_3$). \textbf{C} Projection along the approximate attractor. \textbf{D} $G(X=0, \Delta t)$ \textit{vs}. $\Delta t$ as measured from simulations (black). Error bars: standard deviation of the mean (blue). Red: fit to an OU process. Here $N=1.5 \times 10^{5}$ (compare with Fig.~5C in the main text). \textbf{E} Diffusion coefficient as a function of $N$, with fit to $\propto 1/N$ dependence in red (compare with Fig.~5D in the main text). \textbf{F} Absolute distance of the activity from the symmetry plane $X=0$, averaged over time and over connectivity instances, plotted \textit{vs}. $N$.
Red error bars represent the standard deviation of the mean. 
In all panels $K=1000$ and $N=1.5\times 10^5$. Other parameters are as in Fig.~2.
}
\label{fig:all2all_nonsym}
\end{figure}

\begin{figure}
\centering
\medskip 
\includegraphics[scale=1.0]{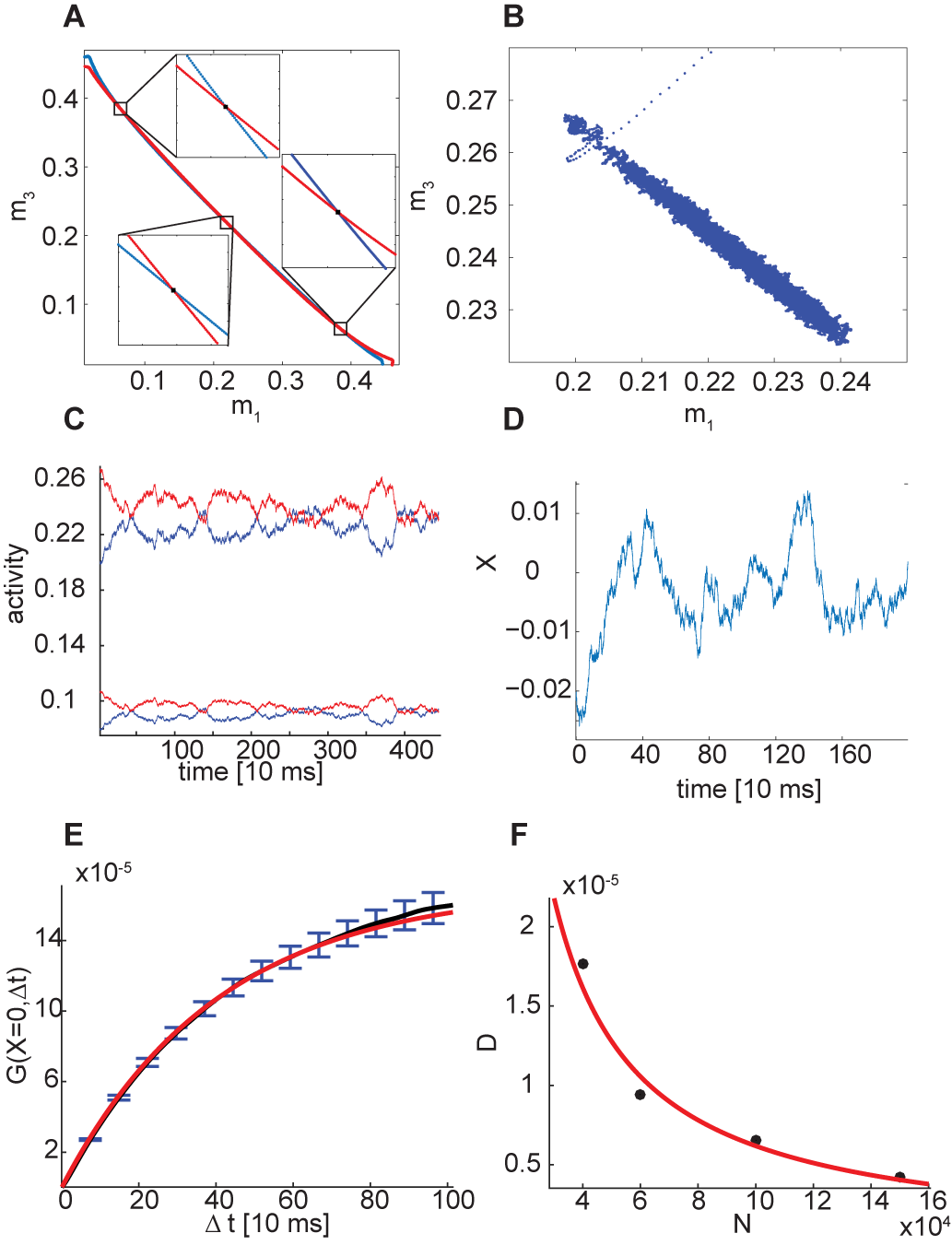}
\vspace{2mm}
\caption{\textbf{ Network with random and sparse inhibitory connections between the sub-networks.} \textbf{A} Projections of the nullclines $\dot{m}_1=0$ (blue) and $\dot{m}_3=0$ (red) on the $m_1 - m_3$ plane, based on Eqs.~\ref{full_sys} and \ref{alpha_random}. Here $K=1000$, $\tilde{J}=1.8$. Insets show a schematic illustration of the nullclines near the fixed points, in which the angle between the lines is amplified for clarity. \textbf{B} Population activities projected onto the $m_1 - m_3$ plane. \textbf{C} Mean activities of the four populations: blue for one subnetwork and red for the other. The higher activities are those of the excitatory populations ($m_1$ and $m_3$). \textbf{D} Projection along the approximate attractor. \textbf{E} $G(X=0, \Delta t)$ 
\textit{vs}. $\Delta t$ as measured from simulations (black, with std of the mean errorbars in blue) and a fit to an OU process (red). (Compare with Fig.~5C in the main text.) \textbf{F} Diffusion coefficient as a function of $N$. Red: fit to $\propto 1/N$ dependence
(compare with Fig.~5D in the main text).
 Here $K=1000$, $N=1.5\times 10^5$,  $\tilde{J}\approx 1.76$, and all other parameters are as in Fig.~2.
 }
\label{fig:not_all2all}
\end{figure}

\begin{figure}
\centering
\medskip
\includegraphics[scale=1.0]{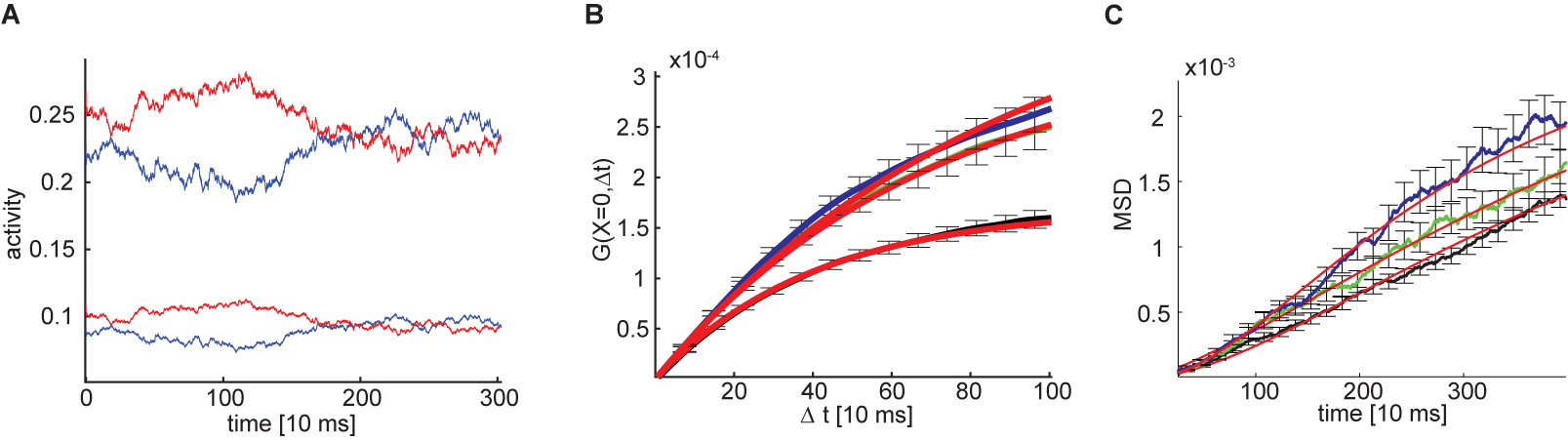}
\vspace{2mm}
\caption{\textbf{ Effects of correlated input noise.} Results from simulations, for a network in which dynamical noise is added to the input $E_0$. 
The noise has a correlation time of $3\tau$, which ensures that the correlation across neurons is not
averaged out due to the asynchronous updating. The noise is described by an OU process:
$\tau_{noise} \dot{\xi}=-\xi+\sigma_{noise} \eta(t)$,
where $\tau_{noise}=30$ ms, and $\eta(t)$ is a gaussian white noise, and the value of $\sigma_{noise}$ was varied to control the noise amplitude.  \textbf{A} Mean activities of the four populations for $\sigma=E_0/3$: blue for one subnetwork and red for the other. The higher activities are those of the excitatory populations ($m_1$ and $m_3$). \textbf{B} $G(X=0,\Delta t)$ for $\sigma_{noise}=0$ (black), $\sigma_{noise}=E_0/30$ (green) and $\sigma_{noise}=E_0/3$ (blue). Error bars represent the standard deviation of the mean. Red: fits to an OU process.
\textbf{C} Mean square displacement (MSD) of the location along the line for initial location $X(0)=0.05$. Colors are the same as in \textbf{B}.
In this figure $K=500$, $N=1.5\times 10^5$, $\tilde{J}\simeq 1.77$.
}
\label{fig:MSD_noise}
\end{figure}

\end{document}